%% file: main.tex
\newcommand{\docstyle}{1} 
\ifnum\docstyle=0
\documentclass[final,5p,times,twocolumn]{elsarticle}
\else
\documentclass[10pt,journal,compsoc]{IEEEtran}
\fi

\usepackage{acronym}
\usepackage{graphicx}
\usepackage{tabularx}
\usepackage{enumerate}
\usepackage{color}
\usepackage{amsmath}
\usepackage{paralist}
\usepackage{threeparttable}
\usepackage{booktabs}
\usepackage{colortbl,hhline}
\usepackage{boldline}
\usepackage{colortbl}
\usepackage{array}
\usepackage{adjustbox}
\usepackage{multirow,tabu}
\usepackage{algorithm,algpseudocode}
\usepackage{tikz,pgfplots}
\usetikzlibrary{arrows}
\usepackage{pgfplotstable}
\usepgfplotslibrary{units,groupplots}
\usepackage{wrapfig}
\usepackage{varwidth,xcolor}
\usepackage{soul}
\usepackage{subcaption}
\usepackage{stfloats}
\usepackage{fancyhdr}

\newcommand{\highlightnewtext}{0}
\ifnum\highlightnewtext=1
\newcommand{\revised}[1]{{\leavevmode\color{blue}{#1}}}
\else
\newcommand{\revised}[1]{{\leavevmode\color{black}{#1}}}
\fi

\newcommand{\mytinysize}{\fontsize{6}{7}\selectfont}
\pgfplotsset{
	compat = newest,
	tick label style={font=\sffamily\scriptsize},
	label style={font=\sffamily\scriptsize},
	legend style={font=\sffamily\mytinysize\raggedleft},
	legend cell align=left,
	grid style={dotted,gray}
}

\newcolumntype{?}{!{\vrule width 0.8pt}}

\definecolor{mygray}{RGB}{220,220,220}
\definecolor{skyblue}{RGB}{86,180,233}
\definecolor{bluish-green}{RGB}{0,158,115}
\definecolor{myblue}{RGB}{86,114,178}
\definecolor{vermilion}{RGB}{213,94,0}
\definecolor{reddish-purple}{RGB}{204,121,167}
\definecolor{deeplilac}{rgb}{0.6, 0.33, 0.73}
\definecolor{darkgreen}{rgb}{0.0, 0.2, 0.13}

\newlength{\Oldarrayrulewidth}

\graphicspath{{artworks/}{results/}}
\DeclareGraphicsExtensions{.pdf,.jpeg,.png}
\ifnum\docstyle=0
	\journal{Computers \& Security}
\fi

\begin{document}

\ifnum\docstyle=0
\title{FLAD: Adaptive Federated Learning for DDoS Attack Detection}
\else
\title{FLAD: Adaptive Federated Learning for\\DDoS Attack Detection}
\fi

\author{Roberto Doriguzzi-Corin, Domenico Siracusa\\
\small\textit{Cybersecurity Centre, Fondazione Bruno Kessler, Italy}
}

\input{acronyms}

\ifnum\docstyle=0
\input{abstract}

\maketitle
\else
\maketitle
\input{abstract}
\fi

\thispagestyle{fancy}
\renewcommand{\headrulewidth}{0pt}
\chead{\scriptsize This is the author's version of an article that has been published in Elsevier Computers \& Security. Changes were made to\\this version by the publisher prior to publication. The final version of record is available at {\color{blue}{https://doi.org/10.1016/j.cose.2023.103597}}. \\ The source code associated with this project is available at {\color{blue}{https://github.com/doriguzzi/flad-federated-learning-ddos}}.}
\cfoot{\scriptsize Copyright (c) 2023 Elsevier. Personal use is permitted. For any other purposes, permission must be obtained from Elsevier by contacting the Permissions Helpdesk Support Center {\color{blue}{https://service.elsevier.com/app/contact/supporthub/permissions-helpdesk/}}.}

\input{charts}
\input{introduction}
\input{motivation}
\input{related}
\input{background}
\input{methodology}
\input{dataset}
\input{setup}
\input{evaluation}
\input{discussion}
\input{conclusions}
\input{ack}

\bibliographystyle{IEEEtran} 
\bibliography{bibliography}

\end{document}

%% file: acronyms.tex
\acrodef{acl}[ACL]{Access Control List}
\acrodef{ai}[AI]{Artificial Intelligence}
\acrodef{ahd}[AHD]{Abnormal Health Detection}
\acrodef{ann}[ANN]{Artificial Neural Network}
\acrodef{api}[API]{Application Programming Interface}
\acrodef{bow}[BoW]{Bag-of-Words}
\acrodef{cnn}[CNN]{Convolutional Neural Network}
\acrodef{cpe}[CPE]{Customer Premise Equipment}
\acrodef{dl}[DL]{Deep Learning}
\acrodef{dlp}[DLP]{Data Loss/Leakage Prevention}
\acrodef{dpi}[DPI]{Deep Packet Inspection}
\acrodef{dnn}[DNN]{Deep Neural Network}
\acrodef{dns}[DNS]{Domain Name System}
\acrodef{dos}[DoS]{Denial of Service}
\acrodef{ddos}[DDoS]{Distributed Denial of Service}
\acrodef{ebpf}[eBPF]{extended Berkeley Packet Filter}
\acrodef{ewma}[EWMA]{Exponential Weighted Moving Average}
\acrodef{fl}[FL]{Federated Learning}
\acrodef{flddos}[\textsc{FLDDoS}]{Federated Learning DDoS}
\acrodef{ourtool}[\textsc{FLAD}]{adaptive Federated Learning Approach to DDoS attack detection}
\acrodef{foss}[FOSS]{Free and Open-Source Software}
\acrodef{fpr}[FPR]{False Positive Rate}
\acrodef{fvg}[\textsc{FedAvg}]{Federated Averaging}
\acrodef{gdpr}[GDPR]{General Data Protection Regulation}
\acrodef{gpu}[GPU]{Graphics Processing Unit}
\acrodef{ha}[HA]{Hardware Appliance}
\acrodef{hids}[HIDS]{Host-based Intrusion Detection System}
\acrodef{ics}[ICS]{Industrial Control System}
\acrodef{ids}[IDS]{Intrusion Detection System}
\acrodef{iid}[i.i.d.]{independent and identically distributed}
\acrodef{ilp}[ILP]{Integer Linear Programming}
\acrodef{iot}[IoT]{Internet of Things}
\acrodef{isp}[ISP]{Internet Service Provider}
\acrodef{ips}[IPS]{Intrusion Prevention System}
\acrodef{jsd}[JSD]{Jensen-Shannon Distance}
\acrodef{ldap}[LDAP]{Lightweight Directory Access Protocol}
\acrodef{lstm}[LSTM]{Long Short-Term Memory}
\acrodef{mano}[NFV MANO]{NFV Management and Orchestration}
\acrodef{mbgd}[MBGD]{Mini-Batch Gradient Descent}
\acrodef{mips}[MIPS]{Millions of Instructions Per Second}
\acrodef{ml}[ML]{Machine Learning}
\acrodef{mlp}[MLP]{Multi-Layer Perceptron}
\acrodef{mssql}[MSSQL]{Microsoft SQL}
\acrodef{nat}[NAT]{Network Address Translation}
\acrodef{netbios}[NetBIOS]{Network Basic Input/Output System}
\acrodef{nic}[NIC]{Network Interface Controller}
\acrodef{nids}[NIDS]{Network Intrusion Detection System}
\acrodef{nf}[NF]{Network Function}
\acrodef{nfv}[NFV]{Network Function Virtualization}
\acrodef{nsc}[NSC]{Network Service Chaining}
\acrodef{ntp}[NTP]{Network Time Protocol}
\acrodef{of}[OF]{OpenFlow}
\acrodef{ood}[o.o.d.]{out-of-distribution}
\acrodef{os}[OS]{Operating System}
\acrodef{pess}[PESS]{Progressive Embedding of Security Services}
\acrodef{pop}[PoP]{Point of Presence}
\acrodef{portmap}[Portmap]{Port Mapper}
\acrodef{ppv}[PPV]{Positive Predictive Value}
\acrodef{ps}[PS]{Port Scanner}
\acrodef{qoe}[QoE]{Quality of Experience}
\acrodef{qos}[QoS]{Quality of Service}
\acrodef{rnn}[RNN]{Recurrent Neural Network}
\acrodef{sdn}[SDN]{Software Defined networking}
\acrodef{sla}[SLA]{Service Level Agreement}
\acrodef{snf}[SNF]{Security Network Function}
\acrodef{snmp}[SNMP]{Simple Network Management Protocol}
\acrodef{ssdp}[SSDP]{Simple Service Discovery Protocol}
\acrodef{svm}[SVM]{Support Vector Machine}
\acrodef{tc}[TC]{Traffic Classifier}
\acrodef{tftp}[TFTP]{Trivial File Transfer Protocol}
\acrodef{tor}[ToR]{Top of Rack}
\acrodef{tpr}[TPR]{True Positive Rate}
\acrodef{tsp}[TSP]{Telecommunication Service Provider}
\acrodef{unb}[UNB]{University of New Brunswick}
\acrodef{vm}[VM]{Virtual Machine}
\acrodef{vne}[VNE]{Virtual Network Embedding}
\acrodef{vnep}[VNEP]{Virtual Network Embedding Problem}
\acrodef{vnf}[VNF]{Virtual Network Function}
\acrodef{vsnf}[VSNF]{Virtual Security Network Function}
\acrodef{vpn}[VPN]{Virtual Private Network}
\acrodef{xdp}[XDP]{eXpress Data Path}
\acrodef{wan}[WAN]{Wide Area Network}
\acrodef{waf}[WAF]{Web Application Firewall}

%% file: abstract.tex
\begin{abstract}
    \acf{fl} has been recently receiving increasing consideration from the cybersecurity community as a way to collaboratively train deep learning models with distributed profiles of cyber threats, with no disclosure of training data. Nevertheless, the adoption of \ac{fl} in cybersecurity is still in its infancy, and a range of practical aspects have not been properly addressed yet. Indeed, the Federated Averaging algorithm at the core of the \ac{fl} concept requires the availability of test data to control the \ac{fl} process. Although this might be feasible in some domains, test network traffic of newly discovered attacks cannot be always shared without disclosing sensitive information.
    In this paper, we address the convergence of the \ac{fl} process in dynamic cybersecurity scenarios, where the trained model must be frequently updated with new recent attack profiles to empower all members of the federation with the latest detection features. To this aim, we propose \acs{ourtool} (\acl{ourtool}), an \ac{fl} solution for cybersecurity applications based on an adaptive mechanism that orchestrates the \ac{fl} process by dynamically assigning more computation to those members whose attacks profiles are harder to learn, without the need of sharing any test data to monitor the performance of the trained model. Using a recent dataset of \acs{ddos} attacks, we demonstrate that \acs{ourtool} outperforms state-of-the-art \ac{fl} algorithms in terms of convergence time and accuracy across a range of unbalanced datasets of heterogeneous \acs{ddos} attacks. We also show the robustness of our approach in a realistic scenario, where we retrain the deep learning model multiple times to introduce the profiles of new attacks on a pre-trained model.
    \end{abstract}
    
    \newcommand{\keywords}{Network Security, Intrusion Detection, Distributed Denial of Service, Federated Learning, Heterogeneous Data}

%% file: charts.tex
\input{results/feature-distribution-PacketLength}

\input{results/results-data-retraining}
\input{results/results-data-full_training-nonIID}
\input{results/results-data-full_training-FLDDoS}
\input{results/results-data-scalability}

\definecolor{flad}{rgb}{1.0, 0.0, 0.0}
\definecolor{fvg1}{rgb}{0.0, 0.75, 1.0}
\definecolor{fvg5}{rgb}{1.0, 0.0, 1.0}
\definecolor{flddos}{rgb}{0.55, 0.71, 0.0}

\newcommand{\chartPacketLengthDistribution}{
	\begin{tikzpicture}
		\begin{axis}[ ybar, 
			legend style={
				at={(0.85,0.95)},
				anchor=north,
				legend columns=1,
				/tikz/every even column/.append style={column sep=0.5cm}
			},
			height=5 cm,
			width=1\linewidth,
			grid = both,
			xlabel={Packet Length (bytes)},
			ylabel={Probability density},
			scaled y ticks=false,
			scaled x ticks=false,
			xmin=0,
			xmax=800,
			xtick={0,100,200,300,400,500,600,700,800},
			xticklabels={0,100,200,300,400,500,600,700,800},
			xtick pos=left,
			ymin=0, ymax=1,
			ytick={0,0.2,0.4,0.6,0.8,1},
			yticklabels={0,0.2,0.4,0.6,0.8,1},
			ytick pos=left,
			enlargelimits=0.05,
			]
			\addplot [cyan,ybar,fill, fill opacity=0.9,  bar width = 7, bar shift = 0] table[x index=0,y index=1] {\FeatureDistributionPacketLength};
			\addplot [deeplilac,ybar,fill, fill opacity=0.9,  bar width = 7, bar shift = -16] table[x index=0,y index=2] {\FeatureDistributionPacketLength};
			\addplot [blue,ybar,fill, fill opacity=0.9,  bar width = 7, bar shift = -8] table[x index=0,y index=6] {\FeatureDistributionPacketLength};
			\addplot [orange,ybar,fill, fill opacity=0.9,  bar width = 7, bar shift = 0] table[x index=0,y index=8] {\FeatureDistributionPacketLength};
			\addplot [red,ybar,fill, fill opacity=0.9,  bar width = 7, bar shift = 0] table[x index=0,y index=9] {\FeatureDistributionPacketLength};
			\addplot [black,ybar,fill, fill opacity=0.9,  bar width = 7, bar shift = 16] table[x index=0,y index=10] {\FeatureDistributionPacketLength};
			\addplot [green,ybar,fill, fill opacity=0.9,  bar width = 7, bar shift = 24] table[x index=0,y index=13] {\FeatureDistributionPacketLength};
			\legend{WebDDoS, LDAP, NTP, SSDP, Syn, TFTP, MSSQL}
		\end{axis}
	\end{tikzpicture}
}

\newcommand{\chartFScoreFullTraining}{
	\begin{tikzpicture}
		\begin{axis}[  
			legend pos=south east,
			legend columns=1,
			height=4.5 cm,
			width=1\linewidth,
			grid = both,
			xlabel={Rounds},
			ylabel={F1 Score},
			scaled y ticks=false,
			scaled x ticks=false,
			xmin=1,
			xmax=62,
			xtick={1,10,20,30,40,50,60},
			xticklabels={1,10,20,30,40,50,60},
			xtick pos=left,
			ymin=0, ymax=1,
			ytick={0,0.1,0.2,0.3,0.4,0.5,0.6,0.7,0.8,0.9,1},
			yticklabels={0,0.1,0.2,0.3,0.4,0.5,0.6,0.7,0.8,0.9,1},
			ytick pos=left,
			enlargelimits=0.02,
			]
			\addplot [color=flad,style={very thick}] table[x index=0,y index=1] {\FScoreFullnonIID};
			\addplot [color=fvg1,style={very thick},dashed] table[x index=0,y index=2] {\FScoreFullnonIID};
			\addplot [color=fvg5,style={very thick},densely dotted] table[x index=0,y index=3] {\FScoreFullnonIID};
			\addplot [color=flddos,style={very thick},dashdotted] table[x index=0,y index=4] {\FScoreFullnonIID};
			\legend{\textbf{\ac{ourtool}}: F1 Score: 0.96; Time: 446 sec, \textbf{\ac{fvg} (E=1)}: F1 Score: 0.83; Time: 2054 sec , \textbf{\ac{fvg} (E=5)}: F1 Score: 0.89; Time: 9842 sec, \textbf{FLDDoS}: F1 Score: 0.95; Time: 12004 sec}
		\end{axis}
	\end{tikzpicture}
}

\newcommand{\chartFScoreWebDDoS}{
	\begin{tikzpicture}
        \begin{axis}[  
			legend style={at={(1.1,1.05)},anchor=south},
            legend columns=4,
            height=5 cm,
            width=1\linewidth,
            grid = both,
            xlabel={Rounds},
            ylabel={F1 Score},
            scaled y ticks=false,
            scaled x ticks=false,
            xmin=1,
            xmax=62,
			xtick={1,10,20,30,40,50,60},
            xticklabels={1,10,20,30,40,50,60},
            xtick pos=left,
            ymin=0, ymax=1,
			ytick={0,0.1,0.2,0.3,0.4,0.5,0.6,0.7,0.8,0.9,1},
			yticklabels={0,0.1,0.2,0.3,0.4,0.5,0.6,0.7,0.8,0.9,1},
            ytick pos=left,
            enlargelimits=0.02,
            ]
			\addplot [color=flad,style={very thick}] table[x index=0,y index=1] {\FScoreFullWebDDoSnonIID};
			\addplot [color=fvg1,style={very thick},dashed] table[x index=0,y index=2] {\FScoreFullWebDDoSnonIID};
			\addplot [color=fvg5,style={very thick},densely dotted] table[x index=0,y index=3] {\FScoreFullWebDDoSnonIID};
			\addplot [color=flddos,style={very thick},dashdotted] table[x index=0,y index=4] {\FScoreFullWebDDoSnonIID};
            \node[draw,align=left,font=\sffamily\scriptsize] at (55,0.2) {WebDDoS};
			\legend{\textbf{\ac{ourtool}}, \textbf{\ac{fvg} (E=1)}, \textbf{\ac{fvg} (E=5)}, \textbf{\ac{flddos}}}
        \end{axis}
    \end{tikzpicture}
}

\newcommand{\chartFScoreSyn}{
	\begin{tikzpicture}
        \begin{axis}[  
            height=5 cm,
            width=1\linewidth,
            grid = both,
            xlabel={Rounds},
            ylabel={F1 Score},
            scaled y ticks=false,
            scaled x ticks=false,
            xmin=1,
            xmax=62,
			xtick={1,10,20,30,40,50,60},
			xticklabels={1,10,20,30,40,50,60},
            xtick pos=left,
            ymin=0, ymax=1,
			ytick={0,0.1,0.2,0.3,0.4,0.5,0.6,0.7,0.8,0.9,1},
			yticklabels={0,0.1,0.2,0.3,0.4,0.5,0.6,0.7,0.8,0.9,1},
            ytick pos=left,
            enlargelimits=0.02,
            ]
			\addplot [color=flad,style={very thick}] table[x index=0,y index=1] {\FScoreFullSynnonIID};
			\addplot [color=fvg1,style={very thick},dashed] table[x index=0,y index=2] {\FScoreFullSynnonIID};
			\addplot [color=fvg5,style={very thick},densely dotted] table[x index=0,y index=3] {\FScoreFullSynnonIID};
			\addplot [color=flddos,style={very thick},dashdotted] table[x index=0,y index=4] {\FScoreFullSynnonIID};
            \node[draw,align=left,font=\sffamily\scriptsize] at (55,0.2) {Syn Flood};
        \end{axis}
    \end{tikzpicture}
}

\newcommand{\chartFScoreFiftyClients}{
	\begin{tikzpicture}
		\begin{axis}[  
			legend pos=south east,
			legend columns=1,
			height=5 cm,
			width=1\linewidth,
			grid = both,
			xlabel={Rounds},
			ylabel={F1 Score},
			scaled y ticks=false,
			scaled x ticks=false,
			xmin=1,
			xmax=45,
			xtick={1,10,20,30,40,50,60,70,80},
			xticklabels={1,10,20,30,40,50,60,70,80},
			xtick pos=left,
			ymin=0, ymax=1,
			ytick={0,0.1,0.2,0.3,0.4,0.5,0.6,0.7,0.8,0.9,1},
			yticklabels={0,0.1,0.2,0.3,0.4,0.5,0.6,0.7,0.8,0.9,1},
			ytick pos=left,
			enlargelimits=0.02,
			]
			\addplot [color=flad,style={very thick}] table[x index=0,y index=1] {\FScoreFullFLDDoS};
			\addplot [color=fvg1,style={very thick},dashed] table[x index=0,y index=2] {\FScoreFullFLDDoS};
			\addplot [color=fvg5,style={very thick},densely dotted] table[x index=0,y index=3] {\FScoreFullFLDDoS};
			\addplot [color=flddos,style={very thick},dashdotted] table[x index=0,y index=4] {\FScoreFullFLDDoS};
			\legend{\textbf{\ac{ourtool}}: F1 Score: 0.9899; Time: 363 sec, \textbf{\ac{fvg} (E=1)}: F1 Score: 0.9255; Time: 1976 sec , \textbf{\ac{fvg} (E=5)}: F1 Score: 0.9339; Time: 9599 sec, \textbf{FLDDoS}: F1 Score: 0.9295; Time: 18319 sec}
		\end{axis}
	\end{tikzpicture}
}

\newcommand{\chartTimeFullTraining}{
	\begin{tikzpicture}
		\begin{axis}[  
			legend pos=north west,
			legend columns=1,
			height=5 cm,
			width=1\linewidth,
			grid = both,
			xlabel={Rounds},
			ylabel={Time (sec)},
			scaled y ticks=false,
			scaled x ticks=false,
			xmin=1,
			xmax=59,
			xtick={1,10,20,30,40,50,60},
			xticklabels={1,10,20,30,40,50,60},
			xtick pos=left,
			ymin=0, ymax=20500,
			ytick={0,5000,10000,15000,20000},
			yticklabels={0,5K,10K,15K,20K},
			ytick pos=left,
			enlargelimits=0.02,
			]
			\addplot [color=red,style={very thick}] table[x index=0,y index=1] {\TimeFullTest};
			\addplot [color=bluish-green,style={very thick},dashed] table[x index=0,y index=2] {\TimeFullTest};
			\addplot [color=blue,style={very thick},densely dotted] table[x index=0,y index=3] {\TimeFullTest};
			\addplot [color=orange,style={very thick},dashdotted] table[x index=0,y index=4] {\TimeFullTest};
			\legend{\ac{ourtool}, \ac{fvg} (E=1), \ac{fvg} (E=5), FLDDoS}
		\end{axis}
	\end{tikzpicture}
}

\newcommand{\chartTimeFiftyClients}{
	\begin{tikzpicture}
		\begin{axis}[  
			legend pos=north west,
			legend columns=1,
			height=5 cm,
			width=1\linewidth,
			grid = both,
			xlabel={Rounds},
			ylabel={Time (sec)},
			scaled y ticks=false,
			scaled x ticks=false,
			xmin=1,
			xmax=52,
			xtick={1,10,20,30,40,50,60},
			xticklabels={1,10,20,30,40,50,60},
			xtick pos=left,
			ymin=0, ymax=20500,
			ytick={0,5000,10000,15000,20000},
			yticklabels={0,5K,10K,15K,20K},
			ytick pos=left,
			enlargelimits=0.02,
			]
			\addplot [color=red,style={very thick}] table[x index=0,y index=1] {\TimeFullFLDDoS};
			\addplot [color=bluish-green,style={very thick},dashed] table[x index=0,y index=2] {\TimeFullFLDDoS};
			\addplot [color=blue,style={very thick},densely dotted] table[x index=0,y index=3] {\TimeFullFLDDoS};
			\addplot [color=orange,style={very thick},dashdotted] table[x index=0,y index=4] {\TimeFullFLDDoS};
			\legend{\ac{ourtool}, \ac{fvg} (E=1), \ac{fvg} (E=5), FLDDoS}
		\end{axis}
	\end{tikzpicture}
}

\newcommand{\chartFScoreProgressive}{
	\begin{tikzpicture}
        \begin{axis}[  
            height=5 cm,
            width=0.9\linewidth,
            grid = both,
            ylabel={F1 Score},
            scaled y ticks=false,
            scaled x ticks=false,
            xmin=2,
            xmax=13,
            xtick={2,3,4,5,6,7,8,9,10,11,12,13},
            xticklabels={},
            xtick pos=left,
            ymin=0.95, ymax=1,
            ytick={0.95,0.96,0.97,0.98,0.99,1},
            yticklabels={0.95,0.96,0.97,0.98,0.99,1},
            ytick pos=left,
            enlargelimits=0.02,
            ]
            \addplot [color=blue,style=thick,mark=*,mark size=1.8] table[x index=0,y index=1] {\FScoreProgressiveFLAD}; \label{fscore_flad_progressive}
        \end{axis}
        \begin{axis}[  
            legend pos=north west,
            legend columns=2,
            height=5 cm,
            width=0.9\linewidth,
            grid = both,
            xlabel={Number of attacks},
            ylabel={F1 Score StdDev},
            scaled y ticks=false,
            scaled x ticks=false,
            xmin=2,
            xmax=13,
            xtick={2,3,4,5,6,7,8,9,10,11,12,13},
            xticklabels={2,3,4,5,6,7,8,9,10,11,12,13},
            xtick pos=left,
            ymin=0, ymax=0.05,
            ytick={0,0.01,0.02,0.03,0.04,0.05},
            yticklabels={0,0.01,0.02,0.03,0.04,0.05},
            ytick pos=right,
            enlargelimits=0.02,
            ]
            \addlegendimage{/pgfplots/refstyle=fscore_flad_progressive}\addlegendentry{F1 Score}
            \addplot [color=orange,style=thick,mark=square*,mark size=1.8] table[x index=0,y index=1] {\FStdProgressiveFLAD}; \label{fstd_flad_progressive}
            \addlegendentry{F1 StdDev}
        \end{axis}
	\end{tikzpicture}
}

\newcommand{\chartScalabilityFLAD}{
	\begin{tikzpicture}
		\begin{axis}[  
			legend pos=north west,
			legend columns=3,
			height=5 cm,
			width=0.94\linewidth,
			grid = both,
			xlabel={Clients},
			ylabel={F1 Score},
			scaled y ticks=false,
			scaled x ticks=false,
			xmin=13,
			xmax=90,
			xtick={13,20,30,40,50,60,70,80,90},
			xticklabels={13,20,30,40,50,60,70,80,90},
			xtick pos=left,
			ymin=0.96, ymax=1,
			ytick={0.94,0.95,0.96,0.965,0.97,0.975,0.98,0.985,0.99,0.995,1},
			yticklabels={0.94,0.95,0.96,0.965,0.97,0.975,0.98,0.985,0.99,0.995,1},
			ytick pos=left,
			enlargelimits=0.02,
			]
			\addplot [color=blue,style= thick,mark=*,mark size=1.8] table[x index=0,y index=1] {\ScalabilityFLAD}; \label{scalability_plot_f1}
			\addlegendentry{F1 Score}
		\end{axis}
		\begin{axis}[  
			legend pos=north west,
			legend columns=3,
			height=5 cm,
			width=0.94\linewidth,
			grid = both,
			xlabel={Clients},
			ylabel={Time (sec)},
			scaled y ticks=false,
			scaled x ticks=false,
			xmin=13,
			xmax=90,
			xtick={13,20,30,40,50,60,70,80,90},
			xticklabels={13,20,30,40,50,60,70,80,90},
			xtick pos=left,
			ymin=100, ymax=900,
			ytick={100,200,300,400,500,600,700,800,900},
			yticklabels={100,200,300,400,500,600,700,800,900},
			ytick pos=right,
			enlargelimits=0.02,
			]
			\addlegendimage{/pgfplots/refstyle=scalability_plot_f1}\addlegendentry{F1 Score}
			\addplot [color=orange,style= thick,mark=square*,mark size=1.8] table[x index=0,y index=3] {\ScalabilityFLAD};
			\addlegendentry{Convergence Time}
		\end{axis}
	\end{tikzpicture}
}

%% file: results/feature-distribution-PacketLength.tex
\pgfplotstableread[header=true]{
	Value WebDDoS LDAP Portmap DNS UDPLag NTP SNMP SSDP Syn TFTP UDP NetBIOS MSSQL
	32.0 0.0 0.006692913385826772 0.029999999999999995 0.0029246775662847006 0.006111249688201546 0.002846405301429874 0.0005985634477254589 0.00011656370206317753 1.0 0.023911257189811012 0.0 0.0 0.0
	46.68 0.8393480791618161 0.003346456692913386 0.0013725490196078432 0.0038356427098815747 0.00037415814417560486 0.007783139496097311 0.0007980845969672786 0.00011656370206317753 0.0 0.0 0.0 0.0 0.0
	61.36 0.004656577415599537 0.0 0.0 0.00023972766936759848 0.0 0.0001334252485045254 0.00039904229848363945 0.0 0.0 0.0 0.0 0.0 0.0
	76.03999999999999 0.0 0.0003937007874015746 0.0005882352941176468 4.7945533873519655e-05 0.0 0.0 0.0001995211492418195 0.0 0.0 0.0 0.0 0.0 0.0001998001998001997
	90.72 0.0 0.0 0.0 9.589106774703931e-05 0.0 0.0 0.0 0.0 0.0 0.0 0.0 0.0 0.0
	105.4 0.0 0.0 0.0 4.79455338735197e-05 0.0 0.004803308946162915 0.0 0.0 0.0 0.0 0.0 0.0 0.0
	120.08 0.0 0.0 0.0 0.0 0.0 0.0 0.0 0.0 0.0 0.0 0.0 0.00019940179461615164 0.0
	134.76 0.0 0.004921259842519683 0.0 0.00014383660162055898 0.0 0.0 0.0 0.0 0.0 0.0 0.0 0.0 0.0
	149.44 0.0 0.0 0.0 0.00014383660162055898 0.0 0.0 0.0 0.0 0.0 0.0 0.0 0.0 0.0
	164.12 0.002328288707799766 0.0001968503937007873 0.0005882352941176468 4.7945533873519655e-05 0.0 0.0 0.0 0.0 0.0 0.0 0.0 0.0 0.0
	178.8 0.0 0.0003937007874015754 0.0 0.00023972766936759872 0.0 0.002757455135760194 0.00019952114924181994 0.0 0.0 0.0 0.0 0.0 0.0
	193.48 0.0 0.001181102362204724 0.0007843137254901956 0.00019178213549407862 0.0 0.0 0.0009976057462090977 0.0 0.0 0.0 0.0 0.0009970089730807572 0.0
	208.16 0.0 0.0035433070866141714 0.0035294117647058807 0.0003356187371146376 0.0 0.0 0.000798084596967278 0.0 0.0 0.0 0.0 0.0029910269192422716 0.0
	222.84 0.0 0.0 0.0 4.7945533873519756e-05 0.0 0.0 0.0 0.0 0.0 0.0 0.0 0.0 0.0
	237.51999999999998 0.001164144353899883 0.1084645669291338 0.09843137254901957 9.589106774703931e-05 0.0 0.00277969267717761 0.06304868316041497 0.0 0.0 0.0 0.0 0.10049850448654032 0.0
	252.2 0.0 0.7923228346456689 0.7741176470588232 0.00019178213549407862 0.0 0.0 0.5155626496408616 0.0 0.0 0.0 0.0 0.8129611166500496 0.0
	266.88 0.0 0.05196850393700785 0.06313725490196075 0.00014383660162055898 0.0 0.0 0.031723862729449305 0.0005828185103158875 0.0 0.0 0.0 0.056829511465603166 0.0
	281.56 0.0 0.011023622047244089 0.01098039215686274 4.7945533873519655e-05 0.0 0.0 0.005586592178770946 0.0009325096165054196 0.0 0.0 0.00034455036177787973 0.010368893320039877 0.0
	296.24 0.0 0.0 0.0 9.589106774703931e-05 0.0009977550511349458 0.0 0.0 0.011539806504254568 0.0 0.0 0.010566211094521643 0.0 0.0
	310.92 0.0 0.004330708661417321 0.0033333333333333314 9.589106774703931e-05 0.0004988775255674729 0.0029353554670995563 0.0017956903431763757 0.0046625480825271 0.0 0.0 0.005053405306075569 0.003788634097706878 0.0
	325.6 0.0 0.0005905511811023641 0.0013725490196078477 0.00019178213549407938 0.0 0.0 0.0001995211492418203 0.0011656370206317792 0.0 0.0 0.000459400482370508 0.0015952143569292177 0.0
	340.28 0.0 0.0003937007874015746 0.0003921568627450978 0.00014383660162055898 0.08692940883013216 0.0 0.000399042298483639 0.0982632008392586 0.0 0.0 0.10313540829217865 0.00019940179461615145 0.0
	354.96 0.0 0.0007874015748031492 0.0001960784313725489 0.0 0.2809927662758791 0.0 0.0001995211492418195 0.2797528849516259 0.0 0.0 0.28965200413460424 0.00019940179461615145 0.0
	369.64 0.0 0.0001968503937007873 0.0 4.7945533873519655e-05 0.002244948865053628 0.0 0.0 0.006177876209348406 0.0 0.0 0.0036752038589640498 0.0 0.0
	384.32 0.0 0.0 0.0 0.0003356187371146376 0.10152157645298074 0.0026017923458382425 0.0 0.08812215875976216 0.0 0.0 0.08820489261513721 0.0 0.0
	399.0 0.0 0.0003937007874015746 0.0 4.7945533873519655e-05 0.22175106011474172 0.0 0.0 0.18615223219489443 0.0 0.0 0.19593430573102094 0.0003988035892323029 0.0
	413.68 0.0034924330616996494 0.0 0.0001960784313725489 4.7945533873519655e-05 0.189324020952856 0.0 0.0 0.1820725026226832 0.0 0.0 0.1931779028367979 0.0 0.0035964035964035947
	428.36 0.017462165308498312 0.0 0.0 0.00023972766936759923 0.09366425542529341 0.0 0.0 0.08812215875976251 0.0 0.0 0.08475938899735874 0.0011964107676969134 0.12507492507492549
	443.03999999999996 0.016298020954598362 0.0 0.0 0.00028767320324111795 0.0008730356697430776 4.447508283484176e-05 0.0 0.010024478377433262 0.0 0.0 0.007120707476742847 0.0013958125623130602 0.09130869130869126
	457.71999999999997 0.015133876600698478 0.0 0.0 0.9651915424078243 0.006859565976552753 0.9733149502990945 0.0 0.004196293274274389 0.0 0.0 0.004594004823705063 0.0011964107676969088 0.09370629370629366
	472.4 0.017462165308498246 0.0 0.0 0.00014383660162055898 0.0 0.0 0.0 0.0008159459144422423 0.0 0.0 0.0 0.0015952143569292116 0.11068931068931065
	487.08 0.005820721769499416 0.0 0.0 9.589106774703931e-05 0.0007483162883512094 0.0 0.0 0.0036134747639585013 0.0 0.0 0.0019524520500746518 0.0003988035892323029 0.05174825174825172
	501.76 0.004656577415599514 0.0 0.0 0.0001438366016205584 0.0002494387627837355 0.0 0.0 0.007693204336169683 0.0 0.0 0.007005857356150194 0.00019940179461615067 0.04095904095904078
	516.44 0.0034924330616996624 0.0 0.0 0.00014383660162055952 0.0 0.0 0.0 0.004779111784590295 0.0 0.0 0.0034455036177788102 0.0003988035892323044 0.03136863136863148
	531.12 0.002328288707799775 0.0 0.0001960784313725497 0.00014383660162055952 0.0002494387627837374 0.0 0.0 0.001282200722694957 0.0 0.9760887428101923 0.00034455036177788104 0.0 0.03756243756243769
	545.8 0.002328288707799757 0.0 0.0007843137254901927 4.794553387351947e-05 0.001995510102269884 0.0 0.0 0.000116563702063177 0.0 0.0 0.00011485012059262611 0.0 0.038161838161838
	560.48 0.008149010477299212 0.0 0.0 4.7945533873519844e-05 0.0009977550511349497 0.0 0.0001995211492418203 0.0 0.0 0.0 0.000459400482370508 0.0 0.02557442557442566
	575.16 0.006984866123399271 0.0 0.004705882352941155 0.00019178213549407789 0.0 0.0 0.0 0.0 0.0 0.0 0.0 0.000598205383848452 0.038161838161838
	589.84 0.0093131548311991 0.0 0.0 0.00028767320324111904 0.002619107009229243 0.0 0.0 0.0 0.0 0.0 0.0 0.0005982053838484567 0.03036963036963047
	604.52 0.0034924330616996355 0.0 0.0 0.0 0.000997755051134942 0.0 0.0 0.0 0.0 0.0 0.0 0.000598205383848452 0.029770229770229643
	619.2 0.0011641443538998875 0.0 0.0 0.00014383660162055952 0.0 0.0 0.0 0.0 0.0 0.0 0.0 0.0001994017946161522 0.016383616383616437
	633.88 0.010477299185098987 0.0 0.0 0.00014383660162055952 0.0 0.0 0.0 0.0 0.0 0.0 0.0 0.0001994017946161522 0.03296703296703308
	648.56 0.004656577415599514 0.0 0.001372549019607837 9.589106774703894e-05 0.0 0.0 0.0 0.0 0.0 0.0 0.0 0.0 0.012387612387612332
	663.24 0.0011641443538998875 0.0 0.0 4.7945533873519844e-05 0.0 0.0 0.0 0.0 0.0 0.0 0.0 0.0001994017946161522 0.022577422577422655
	677.92 0.0 0.0 0.0 0.00019178213549407789 0.0 0.0 0.0 0.0 0.0 0.0 0.0 0.0 0.015584415584415515
	692.6 0.0 0.0 0.0 0.0 0.0 0.0 0.0 0.0 0.0 0.0 0.0 0.0 0.01258741258741263
	707.28 0.0011641443538998786 0.0 0.0 0.0 0.0 0.0 0.0 0.0 0.0 0.0 0.0 0.0 0.009790209790209746
	721.96 0.00465657741559955 0.0 0.0 0.00014383660162055952 0.0 0.0 0.0 0.0 0.0 0.0 0.0 0.0 0.008191808191808219
	736.64 0.0 0.0 0.0 4.7945533873519844e-05 0.0 0.0 0.0 0.0 0.0 0.0 0.0 0.0 0.006993006993007016
	751.3199999999999 0.0 0.0 0.0 0.00019178213549407789 0.0 0.0 0.0 0.0 0.0 0.0 0.0 0.0 0.00539460539460537
	766.0 0.0 0.0 0.0 0.00014383660162055952 0.0 0.0 0.0 0.0 0.0 0.0 0.0 0.0 0.004595404595404611
	780.68 0.0 0.0 0.0 0.0001438366016205584 0.0 0.0 0.0 0.0 0.0 0.0 0.0 0.0 0.004995004995004973
	795.36 0.0 0.0 0.0 0.00019178213549407938 0.0 0.0 0.0 0.0 0.0 0.0 0.0 0.0 0.006593406593406616
	810.04 0.0 0.0 0.0 0.0002397276693675974 0.0 0.0 0.0 0.0 0.0 0.0 0.0 0.0 0.004195804195804177
	824.72 0.0 0.0 0.0 0.0 0.0 0.0 0.0 0.0 0.0 0.0 0.0 0.0 0.004595404595404611
	839.4 0.0 0.0 0.0 0.00019178213549407938 0.0 0.0 0.0 0.0 0.0 0.0 0.0 0.0 0.003596403596403608
	854.0799999999999 0.0 0.0 0.0 0.0001438366016205584 0.0 0.0 0.0 0.0 0.0 0.0 0.0 0.0 0.006593406593406565
	868.76 0.0 0.0 0.0 0.00014383660162055952 0.0 0.0 0.0 0.0 0.0 0.0 0.0 0.0001994017946161522 0.0031968031968032076
	883.4399999999999 0.0 0.0 0.0 9.589106774703894e-05 0.0 0.0 0.0 0.0 0.0 0.0 0.0 0.0 0.004395604395604376
	898.12 0.0 0.0 0.0 0.0 0.0 0.0 0.0 0.0 0.0 0.0 0.0 0.0 0.006993006993007016
	912.8 0.0 0.0 0.0 9.589106774703894e-05 0.0 0.0 0.0 0.0 0.0 0.0 0.0 0.0 0.0031968031968031825
	927.48 0.0 0.0 0.0 0.00019178213549407938 0.0 0.0 0.0 0.0 0.0 0.0 0.0 0.0 0.0019980019980020045
	942.16 0.0 0.0 0.0 0.0 0.0 0.0 0.0 0.0 0.0 0.0 0.0 0.0 0.004195804195804177
	956.84 0.0 0.0 0.0 0.00014383660162055952 0.0 0.0 0.0 0.0 0.0 0.0 0.0 0.0 0.0031968031968032076
	971.52 0.0 0.0 0.0 0.00023972766936759923 0.0 0.0 0.0 0.0 0.0 0.0 0.0 0.0 0.001798201798201804
	986.1999999999999 0.0 0.0 0.0 4.794553387351947e-05 0.0 0.0 0.0 0.0 0.0 0.0 0.0 0.0 0.002597402597402586
	1000.88 0.0 0.0 0.0 0.00028767320324111904 0.0 0.0 0.0 0.0 0.0 0.0 0.0 0.0 0.001798201798201804
	1015.56 0.002328288707799757 0.0 0.003921568627450964 4.794553387351947e-05 0.0 0.0 0.00019952114924181875 0.0 0.0 0.0 0.0 0.0 0.0017982017982017906
	1030.24 0.0 0.0 0.0 0.0002397276693675974 0.0 0.0 0.0 0.0 0.0 0.0 0.0 0.0 0.0001998001998001989
	1044.92 0.0 0.0 0.0 4.794553387352022e-05 0.0 0.0 0.0 0.0 0.0 0.0 0.0 0.0 0.0031968031968032324
	1059.6 0.0 0.0 0.0 0.0001438366016205584 0.0 0.0 0.00019952114924181875 0.0 0.0 0.0 0.0 0.0 0.0031968031968031825
	1074.28 0.0 0.0 0.0 4.794553387351947e-05 0.0 0.0 0.0 0.0 0.0 0.0 0.0 0.0 0.002197802197802188
	1088.96 0.0 0.0 0.0 0.0003356187371146415 0.0 0.0 0.0 0.0 0.0 0.0 0.0 0.0 0.0
	1103.6399999999999 0.0 0.0 0.0 9.589106774703894e-05 0.0 0.0 0.0 0.0 0.0 0.0 0.0 0.0 0.0005994005994005967
	1118.32 0.010477299185098907 0.0 0.0 4.794553387351947e-05 0.0 0.0 0.0003990422984836375 0.0 0.0 0.0 0.0 0.0 0.0015984015984015912
	1133.0 0.0 0.0 0.0 0.0002397276693675974 0.0 0.0 0.0 0.0 0.0 0.0 0.0 0.0 0.0007992007992007956
	1147.68 0.0 0.0 0.0 9.589106774704043e-05 0.0 0.0 0.0 0.0 0.0 0.0 0.0 0.0 0.00039960039960040405
	1162.36 0.0 0.0 0.0 0.0001438366016205584 0.0 0.0 0.0 0.0 0.0 0.0 0.0 0.0 0.0
	1177.04 0.0 0.0 0.0 9.589106774703894e-05 0.0 0.0 0.0011971268954509126 0.0 0.0 0.0 0.0 0.0 0.0
	1191.72 0.0 0.0 0.0 9.589106774703894e-05 0.0 0.0 0.00159616919393455 0.000116563702063177 0.0 0.0 0.0 0.0 0.0005994005994005967
	1206.4 0.0 0.0 0.0 0.00019178213549408087 0.0 0.0 0.0021947326416600407 0.0 0.0 0.0 0.0 0.0 0.0017982017982018181
	1221.08 0.0 0.0 0.0 0.0003356187371146363 0.0 0.0 0.008778930566640026 0.000233127404126354 0.0 0.0 0.0 0.0 0.0007992007992007956
	1235.76 0.0 0.0 0.0 0.00019178213549407789 0.0 0.0 0.006983240223463657 0.0023312740412635407 0.0 0.0 0.0 0.0 0.0003996003996003978
	1250.44 0.0 0.0 0.0 0.00023972766936760108 0.0 0.0 0.01815642458100579 0.01095698799393881 0.0 0.0 0.0 0.0 0.0015984015984016162
	1265.12 0.0 0.0 0.0 0.0002876732032411168 0.0 0.0 0.009976057462090938 0.000116563702063177 0.0 0.0 0.0 0.0 0.0017982017982017906
	1279.8 0.0 0.0 0.0 0.00019178213549407789 0.0 0.0 0.005786113328012744 0.000116563702063177 0.0 0.0 0.0 0.0 0.0015984015984015912
	1294.48 0.0 0.0 0.0 9.589106774703894e-05 0.0 0.0 0.004588986432561832 0.0 0.0 0.0 0.0 0.0 0.010989010989010941
	1309.16 0.0 0.0 0.0 0.0 0.0 0.0 0.007980845969672875 0.00023312740412635764 0.0 0.0 0.0 0.0 0.0
	1323.84 0.0 0.0 0.0 0.00019178213549407789 0.0 0.0 0.009976057462090938 0.000116563702063177 0.0 0.0 0.0 0.0 0.0011988011988011934
	1338.52 0.0 0.0 0.0 0.0002876732032411168 0.0 0.0 0.013367916999201857 0.0 0.0 0.0 0.0 0.0 0.0
	1353.2 0.0 0.0 0.0 9.589106774704043e-05 0.0 0.0 0.012370311252992955 0.00034969110618953645 0.0 0.0 0.0 0.0 0.0005994005994006061
	1367.8799999999999 0.0 0.0 0.0 0.0 0.0 0.0 0.01735833998403823 0.000233127404126354 0.0 0.0 0.0 0.0 0.0
	1382.56 0.0 0.0 0.0 4.794553387351947e-05 0.0 0.0 0.0510774142059056 0.0012822007226949471 0.0 0.0 0.0 0.0 0.0
	1397.24 0.0 0.0 0.0 0.0001438366016205584 0.0 0.0 0.07182761372705476 0.0012822007226949471 0.0 0.0 0.0 0.0 0.0015984015984015912
	1411.92 0.0 0.0 0.0 0.00038356427098816174 0.0 0.0 0.048683160415004534 0.0006993822123790729 0.0 0.0 0.0 0.0 0.0007992007992008081
	1426.6 0.0 0.0 0.0 0.00019178213549407789 0.0 0.0 0.018555466879489144 0.00034969110618953103 0.0 0.0 0.0 0.0 0.0
	1441.28 0.0 0.0 0.0 0.00038356427098815577 0.0 0.0 0.015163607342378226 0.000116563702063177 0.0 0.0 0.0 0.0 0.0001998001998001989
	1455.96 0.0 0.0 0.0 0.00019178213549408087 0.0 0.0 0.02753391859537142 0.00023312740412635764 0.0 0.0 0.0 0.0 0.0013986013986014142
	1470.6399999999999 0.0 0.00039370078740157306 0.0 0.0003356187371146363 0.0 0.0 0.015762170790103683 0.0008159459144422392 0.0 0.0 0.0 0.0 0.0
	1485.32 0.0 0.00846456692913382 0.0 0.01443160569592936 0.0 0.0 0.007581803671189113 0.000116563702063177 0.0 0.0 0.0 0.0 0.0009990009990009947
}\FeatureDistributionPacketLength

%% file: results/results-data-retraining.tex
\pgfplotstableread[header=true]{
Clients E=A,S=A E=A,S=inf E=1,S=A E=1,S=inf E=5,S=A E=5,S=inf
2 0.9674569999999999 0.9707800000000001 0.9614559999999999 0.9667389999999999 0.9674569999999999 0.9707800000000001
3 0.9795130000000001 0.9803599999999999 0.973111 0.969479 0.974465 0.9866629999999998
4 0.9840199999999999 0.9849210000000002 0.9801249999999999 0.9701619999999999 0.9895639999999999 0.9912790000000002
5 0.986326 0.9864170000000001 0.9825889999999999 0.979245 0.983775 0.9920159999999999
6 0.9839930000000002 0.9910120000000001 0.978913 0.9833190000000002 0.986397 0.9923660000000002
7 0.9873450000000001 0.9876150000000001 0.984008 0.9849289999999998 0.985266 0.993017
8 0.986794 0.9894400000000001 0.9875039999999998 0.987152 0.9919789999999999 0.992534
9 0.9874090000000001 0.9898469999999999 0.9853719999999999 0.9864940000000002 0.9908849999999999 0.9927400000000001
10 0.9895880000000001 0.9891659999999998 0.984053 0.9867699999999999 0.9907360000000001 0.992576
11 0.9896320000000001 0.9899770000000002 0.9895049999999997 0.9872460000000001 0.9908319999999999 0.991725
12 0.9906840000000001 0.9898370000000002 0.9875310000000003 0.9869730000000001 0.9917100000000001 0.9915389999999998
13 0.9908959999999999 0.9902380000000001 0.988855 0.9870680000000001 0.9912289999999999 0.9913400000000001
}\FScoreProgressiveFLAD

\pgfplotstableread[header=true]{
Clients E=A,S=A E=A,S=inf E=1,S=A E=1,S=inf E=5,S=A E=5,S=inf
2 0.023918000000000002 0.020600999999999998 0.029435 0.024385000000000004 0.023918000000000002 0.020600999999999998
3 0.019849 0.013615999999999998 0.024907000000000002 0.019061 0.018986999999999997 0.012152
4 0.014292000000000003 0.012205 0.01788 0.019900000000000004 0.006657000000000001 0.008161
5 0.013156000000000001 0.011191999999999997 0.015421 0.014595 0.012633 0.005833
6 0.010573 0.006105 0.016886999999999996 0.016109000000000002 0.009984999999999999 0.005404999999999999
7 0.009996999999999997 0.01311 0.015071999999999999 0.021047 0.011845 0.0052179999999999995
8 0.010252 0.010208000000000002 0.010266 0.012506999999999999 0.006723999999999999 0.0077729999999999995
9 0.010235 0.011486000000000001 0.012834000000000002 0.014586999999999998 0.006857 0.00937
10 0.007228 0.010869 0.013656999999999997 0.012587999999999998 0.00722 0.009004000000000002
11 0.008452 0.007023 0.00736 0.010069999999999999 0.008078 0.009334999999999998
12 0.006769 0.009153999999999999 0.011349999999999999 0.016051000000000003 0.0061 0.011337999999999999
13 0.007082 0.007198999999999999 0.011448000000000002 0.019953999999999996 0.006189000000000001 0.012229
}\FStdProgressiveFLAD

%% file: results/results-data-full_training-nonIID.tex
\pgfplotstableread[header=true]{
Round  FLAD  FedAvg_1e  FedAvg_5e  FLDDoS 
1.0 0.46337 0.58279 0.5839 0.56135
2.0 0.46337 0.59322 0.5839 0.58118
3.0 0.46337 0.79733 0.5839 0.69577
4.0 0.50878 0.81674 0.5839 0.73883
5.0 0.51707 0.81674 0.5839 0.74178
6.0 0.5804 0.81674 0.5839 0.81181
7.0 0.6922 0.81674 0.5839 0.81313
8.0 0.72681 0.81674 0.5839 0.81313
9.0 0.7412 0.81674 0.63861 0.81313
10.0 0.7412 0.81674 0.69385 0.81313
11.0 0.74171 0.81674 0.69385 0.81313
12.0 0.74273 0.81674 0.69385 0.81313
13.0 0.74273 0.81674 0.69385 0.81313
14.0 0.74273 0.81674 0.70944 0.81313
15.0 0.74803 0.81674 0.81408 0.81313
16.0 0.74803 0.81674 0.81408 0.81313
17.0 0.74803 0.81674 0.81408 0.84688
18.0 0.75085 0.81674 0.81464 0.84688
19.0 0.8196 0.81674 0.81464 0.85598
20.0 0.8196 0.81674 0.84349 0.85705
21.0 0.8196 0.81674 0.84349 0.85705
22.0 0.85259 0.81674 0.84349 0.85705
23.0 0.88973 0.81674 0.84349 0.85705
24.0 0.88973 0.81674 0.84349 0.85705
25.0 0.88973 0.81674 0.84349 0.85705
26.0 0.91656 0.81674 0.84349 0.85705
27.0 0.95483 0.81674 0.84349 0.85705
28.0 0.9551 0.82291 0.84349 0.85705
29.0 0.9551 0.82291 0.84349 0.85705
30.0 0.9551 0.82291 0.84349 0.86141
31.0 0.9551 0.82291 0.84349 0.86141
32.0 0.9551 0.82291 0.84349 0.92095
33.0 0.9571 0.82291 0.84349 0.92095
34.0 0.9571 0.82291 0.84349 0.92095
35.0 0.9571 0.82291 0.84349 0.92095
36.0 0.96198 0.82291 0.84349 0.92095
37.0 0.96198 0.82291 0.84349 0.92095
38.0 0.96198 0.82291 0.84349 0.92095
39.0 0.96198 0.82291 0.85039 0.92095
40.0 0.96198 0.82291 0.85039 0.92095
41.0 0.96198 0.82291 0.85039 0.92095
42.0 0.96198 0.82291 0.85039 0.92095
43.0 0.96198 0.82291 0.85039 0.92095
44.0 0.96198 0.84305 0.85039 0.92095
45.0 0.96198 0.84305 0.85039 0.92095
46.0 0.96198 0.84305 0.85039 0.92095
47.0 0.96198 0.84305 0.85428 0.92095
48.0 0.96198 0.84305 0.85428 0.92095
49.0 0.96198 0.84305 0.85428 0.92095
50.0 0.96198 0.84305 0.8729 0.92095
51.0 0.96198 0.84305 0.8729 0.92095
52.0 0.96198 0.84305 0.8729 0.92095
53.0 0.96198 0.84305 0.8729 0.93876
54.0 0.96198 0.84305 0.8729 0.93876
55.0 0.96198 0.85308 0.8729 0.93876
56.0 0.96198 0.85308 0.8729 0.93876
57.0 0.96198 0.85308 0.8729 0.93876
58.0 0.96198 0.85308 0.8729 0.93876
59.0 0.96198 0.85308 0.91675 0.93876
60.0 0.96198 0.85308 0.91675 0.93876
61.0 0.96198 0.85308 0.91675 0.93876
62.0 0.96198 0.85308 0.91675 0.93876
}\FScoreFullnonIID

\pgfplotstableread[header=true]{
Round  FLAD  FedAvg_1e  FedAvg_5e  FLDDoS 
1.0 0.0 0.0 0.0 0.0
2.0 0.0 0.0 0.0 0.0
3.0 0.0 0.0 0.0 0.0
4.0 0.0 0.0 0.0 0.0
5.0 0.0 0.0 0.0 0.0
6.0 0.89655 0.0 0.0 0.0
7.0 0.85714 0.0 0.0 0.0
8.0 0.0 0.0 0.0 0.0
9.0 0.85714 0.0 0.0 0.0
10.0 0.85714 0.0 0.0 0.0
11.0 0.89655 0.0 0.0 0.0
12.0 0.85714 0.0 0.0 0.0
13.0 0.85714 0.0 0.0 0.0
14.0 0.85714 0.0 0.0 0.0
15.0 0.89655 0.0 0.0 0.0
16.0 0.89655 0.0 0.0 0.0
17.0 0.89655 0.0 0.0 0.0
18.0 0.89655 0.0 0.0 0.0
19.0 0.89655 0.0 0.0 0.0
20.0 0.89655 0.0 0.0 0.0
21.0 0.89655 0.0 0.0 0.0
22.0 0.89655 0.0 0.0 0.0
23.0 0.89655 0.0 0.0 0.0
24.0 0.89655 0.0 0.0 0.0
25.0 0.89655 0.0 0.0 0.0
26.0 0.89655 0.0 0.0 0.0
27.0 0.89655 0.0 0.0 0.0
28.0 0.89655 0.0 0.0 0.0
29.0 0.89655 0.0 0.0 0.0
30.0 0.89655 0.0 0.0 0.0
31.0 0.89655 0.0 0.0 0.0
32.0 0.89655 0.0 0.0 0.81481
33.0 0.89655 0.0 0.0 0.81481
34.0 0.89655 0.0 0.0 0.81481
35.0 0.89655 0.0 0.0 0.81481
36.0 0.89655 0.0 0.0 0.81481
37.0 0.89655 0.0 0.0 0.81481
38.0 0.89655 0.0 0.0 0.81481
39.0 0.89655 0.0 0.81481 0.81481
40.0 0.89655 0.0 0.81481 0.81481
41.0 0.89655 0.0 0.81481 0.81481
42.0 0.89655 0.0 0.81481 0.81481
43.0 0.89655 0.0 0.81481 0.81481
44.0 0.89655 0.0 0.81481 0.81481
45.0 0.89655 0.0 0.81481 0.81481
46.0 0.89655 0.0 0.81481 0.81481
47.0 0.89655 0.0 0.81481 0.81481
48.0 0.89655 0.0 0.81481 0.81481
49.0 0.89655 0.0 0.81481 0.81481
50.0 0.89655 0.0 0.81481 0.81481
51.0 0.89655 0.0 0.81481 0.81481
52.0 0.89655 0.0 0.81481 0.81481
53.0 0.89655 0.0 0.81481 0.81481
54.0 0.89655 0.0 0.81481 0.81481
55.0 0.89655 0.0 0.81481 0.81481
56.0 0.89655 0.0 0.81481 0.81481
57.0 0.89655 0.0 0.81481 0.81481
58.0 0.89655 0.0 0.81481 0.81481
59.0 0.89655 0.0 0.81481 0.81481
60.0 0.89655 0.0 0.81481 0.81481
61.0 0.89655 0.0 0.81481 0.81481
62.0 0.89655 0.0 0.81481 0.81481
}\FScoreFullWebDDoSnonIID

\pgfplotstableread[header=true]{
Round  FLAD  FedAvg_1e  FedAvg_5e  FLDDoS 
1.0 0.00043 0.00043 0.00043 0.00043
2.0 0.00043 0.00043 0.00043 0.00043
3.0 0.00043 0.00043 0.00043 0.00043
4.0 0.00043 0.00043 0.00043 0.00043
5.0 0.50417 0.00043 0.00043 0.00043
6.0 0.50357 0.00043 0.00043 0.00043
7.0 0.50813 0.00043 0.00043 0.00043
8.0 0.50935 0.00043 0.00043 0.00043
9.0 0.5095 0.00043 0.00043 0.00043
10.0 0.5095 0.00043 0.00043 0.00043
11.0 0.50989 0.00043 0.00043 0.00043
12.0 0.51109 0.00043 0.00043 0.00043
13.0 0.51109 0.00043 0.00043 0.00043
14.0 0.51109 0.00043 0.00043 0.00043
15.0 0.51181 0.00043 0.00043 0.00043
16.0 0.51181 0.00043 0.00043 0.00043
17.0 0.51181 0.00043 0.00043 0.50401
18.0 0.51197 0.00043 0.00043 0.50401
19.0 0.51268 0.00043 0.00043 0.50401
20.0 0.51268 0.00043 0.50401 0.50401
21.0 0.51268 0.00043 0.50401 0.50401
22.0 0.98806 0.00043 0.50401 0.50401
23.0 0.51276 0.00043 0.50401 0.50401
24.0 0.51276 0.00043 0.50401 0.50401
25.0 0.51276 0.00043 0.50401 0.50401
26.0 0.51307 0.00043 0.50401 0.50401
27.0 0.98928 0.00043 0.50401 0.50401
28.0 0.98873 0.00043 0.50401 0.50401
29.0 0.98873 0.00043 0.50401 0.50401
30.0 0.98873 0.00043 0.50401 0.50903
31.0 0.98873 0.00043 0.50401 0.50903
32.0 0.98873 0.00043 0.50401 0.51094
33.0 0.98668 0.00043 0.50401 0.51094
34.0 0.98668 0.00043 0.50401 0.51094
35.0 0.98668 0.00043 0.50401 0.51094
36.0 0.98796 0.00043 0.50401 0.51094
37.0 0.98796 0.00043 0.50401 0.51094
38.0 0.98796 0.00043 0.50401 0.51094
39.0 0.98796 0.00043 0.5225 0.51094
40.0 0.98796 0.00043 0.5225 0.51094
41.0 0.98796 0.00043 0.5225 0.51094
42.0 0.98796 0.00043 0.5225 0.51094
43.0 0.98796 0.00043 0.5225 0.51094
44.0 0.98796 0.50401 0.5225 0.51094
45.0 0.98796 0.50401 0.5225 0.51094
46.0 0.98796 0.50401 0.5225 0.51094
47.0 0.98796 0.50401 0.5126 0.51094
48.0 0.98796 0.50401 0.5126 0.51094
49.0 0.98796 0.50401 0.5126 0.51094
50.0 0.98796 0.50401 0.52483 0.51094
51.0 0.98796 0.50401 0.52483 0.51094
52.0 0.98796 0.50401 0.52483 0.51094
53.0 0.98796 0.50401 0.52483 0.51222
54.0 0.98796 0.50401 0.52483 0.51222
55.0 0.98796 0.50401 0.52483 0.51222
56.0 0.98796 0.50401 0.52483 0.51222
57.0 0.98796 0.50401 0.52483 0.51222
58.0 0.98796 0.50401 0.52483 0.51222
59.0 0.98796 0.50401 0.5232 0.51222
60.0 0.98796 0.50401 0.5232 0.51222
61.0 0.98796 0.50401 0.5232 0.51222
62.0 0.98796 0.50401 0.5232 0.51222
}\FScoreFullSynnonIID

\pgfplotstableread[header=true]{
Round  FLAD  FedAvg_1e  FedAvg_5e  FLDDoS 
1.0 22.51 40.64 205.74 242.43
2.0 40.16 80.38 410.02 481.95
3.0 57.95 120.83 614.9 727.88
4.0 75.04 140.94 717.82 847.48
5.0 82.36 181.3 921.4 1087.54
6.0 99.25 220.81 1124.38 1325.77
7.0 121.44 240.96 1227.0 1446.34
8.0 136.99 280.51 1431.68 1689.31
9.0 147.43 320.16 1635.46 1926.9
10.0 168.54 330.64 1687.7 1988.56
11.0 185.46 370.05 1891.88 2224.37
12.0 207.27 409.18 2096.26 2464.33
13.0 228.09 448.45 2300.94 2712.36
14.0 243.48 487.34 2505.07 2937.35
15.0 264.67 526.75 2708.3 3175.31
16.0 285.02 566.07 2913.18 3411.77
17.0 299.87 586.22 3015.6 3531.32
18.0 321.48 625.35 3219.83 3766.27
19.0 341.28 635.69 3272.07 3825.94
20.0 355.55 655.42 3374.89 3944.52
21.0 358.61 694.74 3579.47 4179.84
22.0 378.82 714.66 3682.34 4298.42
23.0 393.74 734.53 3785.26 4417.16
24.0 397.84 773.66 3989.54 4651.84
25.0 400.64 793.53 4091.66 4770.42
26.0 410.57 833.04 4298.6 5006.07
27.0 414.89 871.98 4504.04 5240.64
28.0 415.92 911.34 4708.22 5478.5
29.0 416.94 950.75 4912.65 5718.67
30.0 420.74 990.73 5116.53 5953.51
31.0 421.83 1030.38 5319.36 6194.43
32.0 430.61 1050.35 5423.78 6314.25
33.0 434.26 1089.9 5639.05 6550.6
34.0 435.37 1129.17 5861.25 6786.3
35.0 439.13 1168.25 6071.25 7021.95
36.0 441.44 1207.47 6276.34 7257.76
37.0 442.48 1227.39 6379.76 7376.61
38.0 445.97 1266.38 6585.35 7611.34
39.0 446.99 1286.39 6688.27 7730.56
40.0 448.29 1325.71 6895.01 7964.48
41.0 452.79 1364.89 7111.44 8200.94
42.0 454.06 1404.07 7319.89 8437.45
43.0 458.16 1423.99 7424.11 8556.62
44.0 459.24 1434.7 7475.44 8618.39
45.0 463.06 1474.06 7681.13 8857.38
46.0 465.5 1513.71 7888.83 9098.04
47.0 466.6 1553.03 8100.89 9338.43
48.0 470.54 1592.16 8309.99 9576.02
49.0 471.55 1631.52 8520.9 9813.93
50.0 474.77 1670.6 8731.71 10049.31
51.0 475.83 1681.13 8784.25 10109.3
52.0 479.53 1721.06 8990.94 10347.96
53.0 480.58 1741.17 9093.96 10467.88
54.0 481.58 1780.96 9299.95 10708.21
55.0 485.35 1801.11 9405.08 10827.54
56.0 486.73 1840.85 9618.29 11068.68
57.0 487.81 1880.17 9822.57 11305.35
58.0 491.75 1890.51 9877.12 11366.63
59.0 492.8 1910.48 9982.05 11487.2
60.0 496.92 1949.75 10189.69 11725.43
61.0 498.19 1988.83 10398.24 11961.29
62.0 502.54 2028.05 10608.75 12199.31
}\TimeFullnonIID

%% file: results/results-data-full_training-FLDDoS.tex
\pgfplotstableread[header=true]{
Round  FLAD  FedAvg_1e  FedAvg_5e  FLDDoS 
1.0 0.69787 0.5951 0.64686 0.52805
2.0 0.71741 0.81006 0.64686 0.72811
3.0 0.71906 0.86351 0.64686 0.827
4.0 0.77268 0.86351 0.78014 0.827
5.0 0.78067 0.86351 0.87356 0.827
6.0 0.78547 0.86351 0.87356 0.8549
7.0 0.78595 0.86351 0.87356 0.85958
8.0 0.7873 0.86351 0.87356 0.85958
9.0 0.84409 0.86351 0.87356 0.91362
10.0 0.90731 0.86351 0.87356 0.91362
11.0 0.96336 0.86351 0.87356 0.92741
12.0 0.96899 0.86351 0.87356 0.92741
13.0 0.96899 0.86351 0.91624 0.92741
14.0 0.97453 0.86351 0.91624 0.9284
15.0 0.98281 0.86351 0.91624 0.9284
16.0 0.98281 0.86351 0.91624 0.92952
17.0 0.98281 0.91566 0.91672 0.92952
18.0 0.98281 0.91566 0.91913 0.92952
19.0 0.98998 0.9172 0.91913 0.92952
20.0 0.98998 0.9172 0.91989 0.92952
21.0 0.98998 0.9172 0.91989 0.92952
22.0 0.98998 0.9172 0.933 0.92952
23.0 0.98998 0.9172 0.933 0.92952
24.0 0.98998 0.9172 0.933 0.92952
25.0 0.98998 0.9172 0.933 0.92952
26.0 0.98998 0.9172 0.933 0.92952
27.0 0.98998 0.9191 0.933 0.92952
28.0 0.98998 0.9191 0.933 0.92952
29.0 0.98998 0.9191 0.933 0.92952
30.0 0.98998 0.91965 0.933 0.92952
31.0 0.98998 0.91965 0.933 0.92952
32.0 0.98998 0.91965 0.933 0.92952
33.0 0.98998 0.92371 0.933 0.92952
34.0 0.98998 0.92399 0.933 0.92952
35.0 0.98998 0.92399 0.933 0.92952
36.0 0.98998 0.92399 0.933 0.92952
37.0 0.98998 0.92399 0.933 0.92952
38.0 0.98998 0.92399 0.93388 0.92952
39.0 0.98998 0.92399 0.93388 0.92952
40.0 0.98998 0.92399 0.93388 0.92952
41.0 0.98998 0.92399 0.93388 0.92952
42.0 0.98998 0.92399 0.93388 0.92952
43.0 0.98998 0.92399 0.93388 0.92952
44.0 0.98998 0.92399 0.93388 0.92952
45.0 0.98998 0.9255 0.93388 0.92952
}\FScoreFullFLDDoS

\pgfplotstableread[header=true]{
Round  FLAD  FedAvg_1e  FedAvg_5e  FLDDoS 
1.0 21.03 44.85 213.02 461.23
2.0 38.02 88.91 426.14 693.19
3.0 54.77 133.42 639.53 1155.34
4.0 71.62 178.21 853.39 1619.32
5.0 88.56 222.78 1066.51 2082.21
6.0 106.18 269.05 1280.0 2546.3
7.0 122.93 313.11 1492.91 3008.39
8.0 139.01 357.56 1706.93 3470.19
9.0 154.94 400.77 1919.74 3702.32
10.0 175.06 444.94 2133.18 4167.84
11.0 194.22 488.43 2346.83 4399.97
12.0 195.95 532.49 2559.69 4864.58
13.0 199.13 576.72 2771.97 5096.14
14.0 201.2 619.98 2985.83 5559.89
15.0 202.98 663.98 3199.01 6022.09
16.0 204.14 707.41 3412.76 6484.58
17.0 207.12 751.64 3626.51 6715.57
18.0 222.18 795.53 3839.21 7177.54
19.0 228.15 839.64 4052.07 7639.51
20.0 229.64 883.53 4265.09 8102.51
21.0 233.06 927.36 4477.79 8565.06
22.0 248.46 971.25 4691.28 9028.12
23.0 250.72 1014.91 4904.25 9491.53
24.0 252.21 1058.0 5117.27 9722.69
25.0 255.63 1101.66 5331.13 10185.69
26.0 269.49 1144.98 5543.73 10648.98
27.0 271.27 1189.09 5757.33 11110.72
28.0 272.28 1233.03 5971.03 11573.61
29.0 275.12 1276.86 6183.89 12039.42
30.0 285.47 1320.69 6397.28 12504.54
31.0 286.48 1364.52 6611.24 12969.2
32.0 287.88 1408.29 6824.52 13435.87
33.0 290.24 1452.23 7038.33 13669.72
34.0 309.83 1495.32 7252.56 13902.77
35.0 312.43 1538.47 7465.37 14137.31
36.0 313.44 1581.56 7678.76 14604.27
37.0 314.45 1625.39 7892.25 15072.14
38.0 317.58 1669.16 8106.06 15539.27
39.0 331.87 1713.05 8319.92 15773.06
40.0 334.23 1756.2 8532.73 16239.67
41.0 336.59 1800.03 8746.33 16706.8
42.0 339.43 1843.92 8959.72 17174.1
43.0 355.26 1887.98 9172.74 17633.15
44.0 361.18 1931.7 9386.18 18086.19
45.0 362.82 1975.93 9598.88 18319.81
}\TimeFullFLDDoS

%% file: results/results-data-scalability.tex
\pgfplotstableread[header=true, col sep = semicolon]{
Clients;  F1;  F1_std;  Time;  RoundTime;  Rounds; 
13;  0.9667;  0.0369;  617.56;  9.08;  68; 
20;  0.9850;  0.0235;  504.61;  6.73;  75; 
30;  0.9895;  0.0169;  431.12;  4.90;  88; 
40;  0.9905;  0.0184;  460.49;  5.17;  89; 
50;  0.9871;  0.0279;  362.22;  5.17;  70; 
60;  0.9876;  0.0269;  379.47;  4.63;  82; 
70;  0.9912;  0.0157;  360.56;  4.62;  78; 
80;  0.9869;  0.0428;  297.59;  4.96;  60; 
90;  0.9900;  0.0240;  291.63;  4.56;  64; 
}\ScalabilityFLAD

%% file: introduction.tex
\section{Introduction} \label{sec:introduction} 

As the number and complexity of cybersecurity attacks increase at a tremendous pace on a daily basis \cite{purplesec_report}, defenders are in need to find more effective protection measures that rely on machine intelligence. To this account, a recent trend in information security is the adoption of solutions based on \acp{ann} to analyse network traffic and the behaviour of software running on computers to identify possible compromised systems or unauthorised access attempts \cite{berman2019survey,salloum2020machine}. Compared to traditional signature-based and anomaly-based approaches, \ac{ann}-based threat detection methods are more resilient to variations in attack patterns and are not constrained by the requirement to define thresholds for attack detection. However, training and updating an \ac{ann} model for effective threat detection is a non-trivial task, due to the complexity and variability of emerging attacks and the lack of data with relevant and up-to-date attack profiles, especially when dealing with zero-day vulnerabilities.

Collaborative learning is a recent approach that addresses the challenges associated with data and \ac{ann} model updates. It enables multiple independent parties to train and update their \ac{ids} by sharing information on recent attack profiles. In this scenario, a security provider could offer an \ac{ids} trained on incidents experienced by all its customers, ensuring a service that is continuously updated with the latest attacks. Collaborative learning techniques have started to gain attention in recent years, when McMahan et al. \cite{mcmahan2017communication} presented the so-called \acf{fl}, a distributed training approach with focus on the privacy of the individual participants in the \ac{fl} process. \ac{fl} relies on a set of participants (also called clients) that train the model on their local data, and on a central server that aggregates \ac{ann} model parameters collected from clients and distributes the aggregated model back to clients for further training sessions. This sequence of operations is executed multiple times (federated training rounds) with no exchange of clients' private training data, until a target convergence level is reached. 

The application of \ac{fl} in cyber security for intrusion detection has been explored in previous research \cite{flddos,zhang2021flddos,tian2021lightweight}. However, previous works rely on \ac{fvg}, the \ac{fl} mechanism introduced by McMahan et al., which necessitates a representative test set available at the server side to control the training process. We argue that this approach poses a data privacy issue and may restrict the applicability of \ac{fl} in scenarios where only a subset of data classes can be tested by the server. It is reasonable to assume that network data containing recent cyber incidents against one or more clients may include sensitive information that cannot be shared with the server for testing purposes. Consequently, in such cases, the server would not have the ability to assess the performance of the aggregated model using the latest attack traffic. Furthermore, achieving convergence in the \ac{fl} process can present challenges due to several factors. These include the presence of non-\aclu{iid} (non-\ac{iid}) data across clients, as well as unbalanced datasets, which are common in network anomaly detection. Slow convergence can hinder the ability to promptly update the IDS service in response to attacks targeted at specific clients within the federation. While some of these issues have been addressed to some extent in previous works, their effectiveness remains uncertain, as outlined in the subsequent sections.

In this paper, we propose a novel \ac{ourtool}, in which the server verifies the classification accuracy of the global model on clients' validation sets with no exchange of training or validation data, granting that the model is learning from all clients' data and allowing to implement an effective early-stopping regularisation strategy. \ac{ourtool} is conceived to apply \ac{fl} in the cybersecurity domain, where we assume that no attack data will be shared at any time between the clients (e.g., customers of an \ac{ids} service) and the server (e.g., provider of the service). We tackle the convergence of the federated learning process in the context of \ac{ddos} attack detection, with focus on the trade-off between convergence time and accuracy of the merged model in segregating benign network traffic from a range of different \ac{ddos} attack types. We consider a dynamic scenario, where clients are targeted by zero-day \ac{ddos} attacks, and where the global model must be updated with new information as soon as possible to empower all participants with the latest detection features.

The high-level idea behind \ac{ourtool} is to involve in a training round only those clients that do not obtain sufficiently good results on their local validation sets with the current global model. For such clients, the amount of computation (number of training epochs and gradient descent steps/epoch) is determined based on their relative accuracy on their validation sets.
Note that, the accuracy score is computed by clients on their validation sets and communicated to the server upon request. Hence, no exchange of sensitive data between server and clients is involved. Compared to \ac{fvg}, \ac{ourtool} introduces a negligible traffic overhead between clients and server, without disclosing clients' sensitive data, even for testing purposes.

We evaluate \ac{ourtool} in a worst-case scenario, where the \ac{ddos} attack data among the clients is unbalanced and non-\ac{iid}. We compare \ac{ourtool} against \ac{fvg} and FLDDoS \cite{flddos}, a state-of-the-art \ac{ddos} attack detection tool that builds upon \ac{fvg} and is designed to address the issues associated with non-\ac{iid} data.
We demonstrate that \ac{ourtool} improves \ac{fvg} and FLDDoS in terms of training time, number of training rounds/client and classification accuracy on unseen traffic data. That is, our approach allows for a faster global model update, requires less computation on clients, and ensures a high accuracy on all \ac{ddos} attack types.

The main contributions of this work are the following:    
\begin{itemize}
    \item  An analysis of the limitations of the \ac{fvg} algorithm in cybersecurity applications with unbalanced and non-\ac{iid} data.
    \item \ac{ourtool}, a novel adaptive mechanism that addresses the aforementioned limitations by steering the federated training process in terms of client selection and amount of computation for each client.
    \item An extensive evaluation on a recent dataset that compares our approach against the \ac{fvg} algorithm and FLDDoS, demonstrating that \ac{ourtool} is more efficient and outputs aggregated models of higher classification accuracy.
    \item A prototype implementation of \ac{ourtool}, publicly available for testing and use \cite{flad-github}.
\end{itemize}

The remainder of this paper is organised as follows. Section \ref{sec:problem_formulation} presents the \ac{fvg} algorithm and highlights its limitations in training models for cybersecurity applications. Section \ref{sec:related} reviews and discusses the related work. Section \ref{sec:threat_model} provides a threat model analysis. Section \ref{sec:methodology} presents the \ac{ourtool} adaptive federated training for \ac{ddos} attack detection. Sections \ref{sec:dataset} and \ref{sec:setup} detail the dataset and the experimental setup. In Section \ref{sec:evaluation}, \ac{ourtool} is evaluated and compared against state-of-the-art \ac{fl} solutions. Section \ref{sec:discussion} analyses the security risks of FLAD and discusses the available techniques to mitigate them. Finally, the conclusions are given in Section \ref{sec:conclusions}. 

%% file: motivation.tex
\section{Problem Formulation}\label{sec:problem_formulation}
\acf{fl} was introduced in 2017 by McMahan et al. \cite{mcmahan2017communication} as a communication-efficient process for training neural networks on decentralised data. The paper formulates the \ac{fvg} algorithm, which is proposed to optimise the federated learning process in real settings, including non-\ac{iid} and unbalanced datasets. The \ac{fl} process involves a central server and a set of $K$ clients, each with a fixed local dataset. Such a process consists of several \textit{rounds} of federated training during which the server selects a random fraction $F$ of clients (for efficiency reasons) and sends them an \ac{ann} model for local training. The selected clients train the model with local data and send it back to the server, which integrates all the updates with the global model. This process is iterated for several rounds until the desired test-set accuracy is reached. The key aspects in this process are three: the aggregation of local updates, the amount of computation performed at each round and the training-stopping strategy, where the latter assumes the availability of test data at the server location.

The aggregation of clients' updates is based on the \ac{fvg} algorithm, formulated in Equation \ref{eq:fed_avg}, which computes the average of clients' models weighted with the number of local training samples ($n_k$).

\begin{equation}\label{eq:fed_avg}
	w_t\gets\sum_{k=1}^K\frac{n_k}{n}w^k_t
\end{equation}

In Equation \ref{eq:fed_avg}, $n=\sum_{k=1}^K n_k$ is the total number of training samples, while $w^k_t$ represents the set of parameters of client $k$ at round $t$. Please note that the aggregation is always performed using the weights of all $K$ clients, although only a fraction $F$ of them have been updated during round $t$. For all the other clients, the weights of round $t-1$ are used. 

Two main parameters control the amount of computation necessary at each round of the \ac{fl} process. The fraction of clients $F$ that perform local training, and the number of local updates performed by each client $k$, which is computed as $u_k=E\cdot S = E\cdot\frac{n_k}{B}$, where $E$ is the number of training epochs and $S$ is the number of gradient descent steps for each epoch, which depends on the batch size $B$, such that $S=\frac{n_k}{B}$.  McMahan et al. report the outcomes of various experiments on image classification and language modelling tasks in terms of communication rounds with different combinations of $F$, $E$ and $B$, which are kept constant during each experiment.

\subsection{Limitations of \ac{fvg}}\label{sec:fvg_limitations}
In Section \ref{sec:evaluation}, we demonstrate that the \ac{fvg} algorithm, as conceived by McMahan et al., does not satisfy two basic requirements for effective \ac{ddos} attack detection: 
\begin{enumerate}
	\item Short convergence time to reach the target attack detection accuracy, especially in emergency threat situations in which the global model must be quickly distributed to clients upon retraining with recent \ac{ddos} attack information. Indeed, \ac{fvg} assigns the same amount of computation to all the clients selected for a round of training, irrespective of the accuracy level reached by the global model on specific clients' data. This inefficient management can lead to long \ac{fl} training sessions with no substantial gain in accuracy.
	\item Accurate detection of all attack types in realistic conditions, where the detection system must learn from unbalanced and non-\ac{iid} data obtained from heterogeneous \ac{ddos} attack types characterised by different traffic rates and feature distributions. 
	The weighted average of \ac{fvg} gives more importance to \revised{the weights of the clients with large local training sets, to the detriment of the smallest ones. We argue that this strategy could hinder \ac{fvg}'s ability to detect attacks characterised by out-of-distribution features that are available only in small local training sets.}
\end{enumerate}

Furthermore, it should be noted that \ac{fvg} operates under the assumption that some test data is accessible at the server site to verify that a target accuracy of the global model is achieved and stop the training process. We argue that this assumption rarely holds in the cybersecurity domain. For instance, let us consider a scenario where one client contributes with updates related to zero-day attack traffic that is not public at training time. In this case, the only solution for the server to verify that the model has learned the new attack would be to use the client's test set. However, even if we discount the willingness of the client to provide such information, this would require data cleaning (anonymisation) from the client's sensitive information, with the risk of losing IP, transport and application layer features that could be critical for model validation. 

\subsection{Problem statement}\label{sec:problem_statement}
Our problem of \ac{ddos} attack detection in federated environments can be formulated as the maximisation of global model accuracy on unbalanced, non-\ac{iid} data across clients, while minimising total \ac{fl} time.
Based on the previous discussion, the solution must satisfy the following constraints:
\begin{enumerate}
	\item[\textbf{C1}] No training data can be shared among clients or between clients and the server.
	\item[\textbf{C2}] No test data is available at the server location.
\end{enumerate}
The solution to the above problem is challenging due to the competing objectives of maximum accuracy and minimum training time and, on the other hand, the limited data (we assume no data at all) available for the server to assess the performance of the global model during the \ac{fl} process. 

%% file: related.tex
\section{Related work}\label{sec:related}
Implementing a robust and efficient \ac{fl} system is a complex task \cite{kairouz2021advances} that often involves domain-specific tuning and optimization. In cybersecurity, recent works have addressed issues related to non-\ac{iid} and unbalanced data, with a primary focus on performance aspects such as accuracy and convergence time. Nevertheless, it is important to acknowledge that these works heavily depend on the vanilla \ac{fvg} algorithm, thereby inheriting the limitations discussed in Section \ref{sec:fvg_limitations}.
In this section, we provide a comprehensive review and discussion of the current state-of-the-art in \ac{fl} research, with a particular emphasis on challenges specific to the cybersecurity domain. 

\subsection{\acl{fl} in cybersecurity}
In cybersecurity, \ac{fl} methods can be exploited to share attack and anomaly profiles with other parties with no disclosure of private data. 
In this regard, FedOE \cite{pourahmadi2022spotting}, LwResnet \cite{tian2021lightweight}, FIDS \cite{fids} and two tools called FLDDoS \cite{flddos,zhang2021flddos} are recent solutions for \ac{ddos} attack detection evaluated on the CIC-DDoS2019 dataset, the same used to assess the performance of \ac{ourtool} (cf. Section \ref{sec:dataset}).
FedOE is a \ac{fl}-based framwork that resorts on semi-supervised learning to detect \ac{ddos} attacks. Clients share with the server the minimum and maximum anomaly scores obtained on their local datasets. The scores are used to find the optimal threshold that maximises the F1 Score across all the clients. 

LwResnet is a lightweight residual network that has been evaluated in \ac{fl} settings with \ac{fvg}, using a subset of 6 UDP-based attacks out of the 13 attacks available in the CIC-DDoS2019 dataset. The first tool called FLDDoS \cite{flddos} addresses the challenges associated with non-\ac{iid} data by combining \ac{fvg} with a local weighted average conducted by individual clients. The local average incorporates the global model received from the server, along with a local model that is trained exclusively with the client's local data. Given the similarities with our work, in Section \ref{sec:sota_comparison} we will compare \ac{ourtool} against FLDDoS in terms of convergence time and accuracy on non-\ac{iid} data. The authors of the second solution called FLDDoS \cite{zhang2021flddos} propose a methodology that tackles the issue of local data imbalance through data augmentation. In addition, they suggest a two-stage model aggregation approach that helps to reduce the number of federated training rounds. On the other hand, FIDS is proposed to improve the performance of \ac{fvg} on non-\ac{iid} data with feature augmentation. This technique requires sharing a representation of the client's data with the central server. 
Dimolianis et al. \cite{dimolianis2022ddos} focus on collaborative \ac{ddos} attack mitigation using programmable firewalls. In this work, a \ac{mlp} model is trained using \ac{fvg} to avoid sharing private training data. Yin et al. \cite{yin2022trusted} tackle the vulnerability of \ac{fl} to inference and poisoning attacks by applying encryption and blockchain-based reputation techniques to a \ac{fl} framework for \ac{ddos} attack detection.
In a recent paper, Popoola et al.  \cite{popoola2021federated} show the benefits of \ac{fl} in detecting zero-day botnet attacks in \ac{iot} environments. The whole study is focused on the application of \ac{fvg} on traffic generated by infected \ac{iot} devices (including the Mirai \cite{antonakakis2017understanding} botnet) and compares \ac{fvg} against other training approaches, either centralised or distributed.

A common approach to tackle anomaly detection problems consists of training a \ac{ml} model with data collected during normal operations of the monitored environment (i.e., free from anomalies). However, in federated infrastructures, this might require sharing sensitive data among members of the federation. In this regard, \ac{fl} has been exploited in recent works  \cite{nguyen2019diot,liu2020deep,mothukuri2021federated, zhao2019multi, zhao2022semi,friha20232df,friha2022felids} to build privacy-preserving anomaly detection systems for \ac{iot} and computer networks. Although the proposed solutions show high detection accuracy scores, the use of the vanilla \ac{fvg} algorithm makes them prone to the drawbacks presented in Section \ref{sec:problem_formulation}.
Finally, an interesting work by Wang et al. \cite{wang2021non} presents a peer-to-peer variation of \ac{fl} to train a model for anomaly detection in \ac{iot} without the need for a central server. To improve convergence and accuracy on non-\ac{iid} and imbalanced data, clients share synthetic data with neighbours. Nevertheless, a stopping strategy for peer-to-peer training is not discussed.  

In summary, we note that current threat detection solutions focus on performance (accuracy and communication rounds), with no or little attention to practical aspects. On the one hand, the assumption that test data is available at the central server location does not always hold. This is a common limitation of the works above, related to constraint C2 formulated in Section \ref{sec:problem_statement}, in which it is not clear how the central server verifies the performance of the global model with respect to recent attacks.
On the other hand, a few works rely on the vanilla \ac{fvg} algorithm \cite{pourahmadi2022spotting, fids,dimolianis2022ddos, popoola2021federated, nguyen2019diot, zhao2022semi,friha2022felids}, which aggregates the local models using weighted averaging. We will demonstrate in Section \ref{sec:sota_comparison} that such an approach can greatly increase the convergence time on unbalanced non-\ac{iid} attack data. Moreover, in some cases, constraint C1 is not respected (jeopardising clients' privacy), as data-sharing mechanisms are used to correlate non-\ac{iid} features. 

\subsection{Unbalanced and non-\ac{iid} data}
The accuracy of \acp{ann} trained with \ac{fvg} can degrade significantly in scenarios with class imbalance \cite{wang2021addressing} or with non-\ac{iid} data \cite{zhao2018federated}. 
To mitigate the issues of unbalanced data across clients, Duan et al. \cite{duan2020self} propose Astraea, a framework that combines data augmentation with mediators placed between the central server and clients. The role of each mediator is to reschedule the local training of a subset of clients, which are selected based on their data distribution. 
Zhang et al. \cite{zhang2022federated} propose ranking the clients' models using their accuracy on a public test set before selecting the best-performing ones to be aggregated into a global \ac{ids}.
Briggs et al. \cite{briggs2020federated} apply a hierarchical clustering algorithm that uses clients' updates to determine the similarity of their training data. The algorithm returns a set of clusters, each containing a subset of clients with similar data.   
Wang et al. \cite{wang2021addressing} propose a centralised monitoring system to spot class imbalance in the training data. The monitor relies on clients' data (part of it) to estimate the composition of data across classes.
Zhao et al. \cite{zhao2018federated} demonstrate the weight divergence of \ac{fvg} on non-\ac{iid} data and improve the accuracy of the global model with a strategy that relies on sharing training data among clients.
FAVOR \cite{favor} improves the performance of \ac{fvg} on non-\ac{iid} data with a client selection mechanism based on reinforcement learning. An agent, collocated with the \ac{fl} server, is in charge of selecting the clients that perform computation at each round. The agent takes its decisions using a reward function that evaluates the accuracy of the global model on validation data.  

We observe that none of the above approaches respects constraint C2, as all of them assume test data at the server location to assess model convergence. Moreover, the works of Wang and Zhao rely on sharing portions of clients' data, failing to meet constraint C1.

\subsection{Efficient \acl{fl}}
The efficiency of the \ac{fl} process is particularly relevant in the edge computing domain, where nodes possess a limited amount of resources (compared to cloud environments) to devote to critical or latency-sensitive tasks. In the scientific literature, the problem has been tackled from various angles: optimisation of the local training process, reduction of communication overhead, and minimisation of the number of local training rounds assigned to clients.

The approach of Ji et al. \cite{ji2021dynamic} is to progressively decrease the fraction of clients that perform local computation, while reducing the amount of transmitted data by means of a mechanism that masks part of the parameters of local models. 
Sparse Ternary Compression (STC) is a protocol proposed by Sattler et al. \cite{stc} to compress upstream and downstream communications between server and clients. Evaluation results show that STC converges faster than \ac{fvg} on non-\ac{iid} data with lower communication overhead.  
The adaptive mechanism proposed by Wang et al. \cite{wang2019adaptive} and FedSens \cite{zhang2021fedsens} focuses on improving the local training process.  
The former optimises the number of local gradient descent steps $S$ taken by the clients (edge nodes) while minimising resource consumption (e.g., time, energy, etc.) and global loss function. Similar to \ac{ourtool}, this approach relies on the performance (local loss function) of the global model on local datasets to control the number of local training steps $S$. However, formulation and evaluation of the proposed approach focus on application scenarios where the amount of training data is equally distributed across clients, with global loss and model computed using weighted averaging. 
FedSens implements an asynchronous \ac{fl} framework, where each client can choose at which round to perform local computation. The goal is to find the best trade-off between classification accuracy and energy consumption (which is a function of the frequency of local and global updates). 

Among these four works, only the mechanism of Wang et al. \cite{wang2019adaptive} satisfies both constraints C1 and C2. However, it has been designed for balanced settings, in which a common value of gradient descent steps across clients is sufficient to achieve the target objectives of accuracy and efficiency. We demonstrate in Section \ref{sec:sota_comparison} that the weighted averaging adopted in that work prevents the global model to learn small and out-of-distribution attack classes in a reasonable training time. We also show that assigning specific training parameters to clients (based on the performance of the global model in their validation sets) greatly reduces convergence time. 
\\

Although \ac{fl} offers the potential for collaborative training of deep learning models for \ac{ddos} detection, existing approaches are limited by their suitability for real-world implementations, where new attack profiles must be shared with other partners with privacy guarantees. 
In this direction, we present our \ac{fl} approach to \ac{ddos} detection \ac{ourtool}, which respects constraints C1 and C2, while achieving high detection accuracy across all attack types in a reasonable training time. We demonstrate that \ac{ourtool} is robust to model re-training upon the availability of new attack data. To the best of our knowledge, this is the first study in which the latter aspect is analysed and addressed.

%% file: background.tex
\section{Threat model}\label{sec:threat_model}

We consider a scenario in which the federation is composed of a set of clients that might belong to different organisations, plus an additional entity that manages the \ac{fl} process (the central server). We assume that no one in the federation has the willingness/permission to share network traffic data with others. On the other hand, the federation's goal is to enhance the \ac{ddos} detection capabilities of each client' \acp{ids} with attack profiles owned by other members. 

In such a scenario, the clients are vulnerable to zero-day \ac{ddos} attacks at any given moment. To ensure the highest level of security, our system requires the global model to be updated promptly with the latest information, empowering all participants with the most recent detection features available. However, it is important to note that the central server may not always have access to network traffic profiles associated with these new and evolving threats. As a result, verifying the effectiveness of the global model in classifying such attacks becomes a challenge.

We also assume that neither the server nor the clients are malicious, thus they do not try to compromise the global model with poisoned data (e.g., weights obtained with mislabelled samples).
Poisoning attacks can be a serious concern for \acp{ids} that rely on collaborative training techniques for their operation.  Malicious clients can manipulate the training process by providing mislabeled data or specially crafted samples, resulting in a final model that fails to accurately classify certain types of attack traffic. This problem is prevalent in most machine learning-based cybersecurity applications, including FL-trained \ac{ddos} attack detection systems. While poisoning attacks remain a critical concern for \acp{ids}, previous research has already tackled the issue \cite{yin2022trusted,qu2022blockchain,zhang2022secfednids,lai2023two,lycklama2023rofl}, and it is not within the scope of this work.

In this context, the adversary does not belong to the federation and does not have the knowledge to generate adversarial evasion attacks against the global \ac{ann} model \cite{rosenberg2021adversarial}. However, it knows the IP addresses of the victims and how to generate \ac{ddos} attacks using spoofed network packets with the source IP address of the victims.  

%% file: methodology.tex
\section{Methodology}\label{sec:methodology}

\ac{ourtool} enhances \ac{fvg} to solve the problem formulated in Section \ref{sec:problem_statement}. In summary, the clients share with the server the classification score obtained by the global model on their local validation sets. As the server has a full view of the performance of the global model across the clients and their attacks, it can implement a training-stopping strategy that ensures acceptable performance on all the attack types, with no need for test data of all attacks (not always available at the server side, especially in the case of clients experiencing zero-day attacks). 
Additionally, the server uses this information to dynamically tune the computational workload of clients during each training round. This approach aims to accelerate the \ac{fl} process when dealing with \ac{ood} data, or to alleviate the computational burden on clients whose local attack profiles are rapidly learnt.	

\begin{table}[t!]
	\centering
	\caption{Glossary of symbols.}
	\renewcommand{\arraystretch}{1.2}
	\label{tab:notations}
	\begin{adjustbox}{width=1\linewidth}
	\begin{tabular}{|l|l|}
		\hline
		\textit{$w_t$} & Global model at round $t$ \\
		\hline
		\textit{$w^c_t$} & Model trained by client $c$ at round $t$  \\
		\hline
		\textit{$\bar{w}$} & Trained global model  \\
		\hline
		\textit{$C$} & Set of federated clients  \\
		\hline
		\textit{$c\in C$} & A participant (client) in the \ac{fl} process\\
		\hline
		\textit{$c_e$} & Number of epochs assigned to client $c$ \\
		\hline
		\textit{$c_s$} & Number of \acs{mbgd} steps/epoch assigned to client $c$ \\
		\hline
		\textit{$C_t$} & Subset of clients that perform training at round $t$ \\
		\hline
		\textit{$a^c$} & Accuracy score computed by client $c$ on its validation set \\
		\hline
		\textit{$a^\mu$} & Average value of $a^c$ computed over all $c\in C$  \\
		\hline
		\revised{\textit{$T_s^c$}} & \revised{Time taken by client $c$ to complete an \acs{mbgd} step} \\
		\hline
		\revised{\textit{$T_n^c$}} & \revised{Total time taken by the two-way transmission} \\ 
		& \revised{of the global model between server and client $c$}\\
		\hline
		\textit{$e_{min},e_{max}$} & Minimum and maximum local training epochs  \\
		\hline
		\textit{$s_{min},s_{max}$} & Minimum and maximum local training \acs{mbgd} steps  \\
		\hline
	\end{tabular}
	\end{adjustbox}
\end{table}

With the term ``computation'', we refer to the number of training epochs/round ($c_e$) and the number of steps/epoch in \ac{mbgd} ($c_s$). By multiplying these two values, we obtain the total number of \ac{mbgd} steps/round allocated to a client for training the global model on its local dataset. In general, when training a neural network, the weights of the network can be thought of as a point in a high-dimensional space, where each dimension corresponds to an individual weight. 
The objective of the clients' local training process is to find the point that maximises the global model's accuracy on the local validation sets. 	  
In this regard, \ac{ourtool} adopts a personalised training strategy that assigns a specific number of \ac{mbgd} steps per round to each client. This allocation is based on the gap between the current global model's accuracy on the client's validation set and the maximum achievable accuracy. As a result, clients with larger accuracy gaps are required to perform more training steps to converge towards the optimal accuracy point. Conversely, clients with smaller gaps are assigned fewer steps, or even no steps at all, recognising their proximity to the point of maximum accuracy score.

\revised{
Unlike \ac{fvg}, which assigns a fixed amount of computation (values of $c_e$ and $c_s$) to a randomly selected subset of clients at every round, with this approach we aim to save clients' computing resources and to reduce the overall federated training time. This total training time is defined as the cumulative duration of all training rounds until convergence is achieved. In this regard, as the clients train in parallel, the time taken by a round of \ac{fl} depends on the slowest client of the federation, as expressed in Equation \ref{eq:round_time}.

\begin{equation}\label{eq:round_time}
T = \max_{c\in C_t} \{T^c_n+(c_e\cdot c_s)\cdot T^c_s\}
\end{equation}

The round time $T$ is computed as the maximum training time across the subset of clients $C_t$ selected by \ac{ourtool} at round $t$. The time spent by a client in a round of \ac{fl} can be computed as the sum of the network time and computation time. The network time $T^c_n$ is the time necessary for the two-way transmission of the global model between server and client. This time mostly depends on the type and the stability of the communication channel between the two parties. The computation time can be expressed as the sum of the time taken by all \ac{mbgd} steps executed by the client during the \ac{fl} round. The computation time is the result of the multiplication of the number of training epochs/round $c_e$ by the number of \ac{mbgd} steps/epoch $c_s$, by the time $T^c_s$ taken by each step. This time depends on the size of the local training set of the client and the computational power of the client's hardware.
Our intuition is that the convergence performance of \ac{fvg} can be improved by using the clients' classification scores to smartly select the clients at each round and to set per-client and per-round values of $c_e$ and $c_s$. 
By reducing, or even eliminating, the computational workload assigned to clients whose traffic profiles are learned faster, we have the potential to optimise the overall convergence time.
}

We present the details of the federated training process executed by the server with \ac{ourtool} in Algorithms \ref{lst:federated-training} and \ref{lst:update_parameters}, while the local training executed by clients is presented in Algorithm \ref{lst:client-update}. The symbols are defined in Table \ref{tab:notations}.

The pseudo-code in Algorithm \ref{lst:federated-training} describes the main process executed by the server, which orchestrates the operations of the clients.  The algorithm takes as input a global model ($w_0$)  and the set of clients involved in the \ac{fl} process ($C$).  It runs indefinitely until convergence is reached, as controlled by parameter $\textsc{patience}$, which is the number of rounds to continue before exit if no progress is made. 
The federated learning starts with the initialisation of the variables that are used to record the best global model along the process (max accuracy score $a_{max}$) and to implement the early stopping strategy (counter $sc$ keeps track of the rounds with no improvements in average accuracy score $a^{\mu}$). At line \ref{lst:fl-line:emin_emax}, the amount of computation for the clients is set to the maximum values of training epochs and \ac{mbgd} steps. 
The loop at lines \ref{lst:fl-line:update_start}-\ref{lst:fl-line:update_stop} triggers the \textsc{ClientUpdate} methods (Algorithm \ref{lst:client-update}) for a subset of selected clients $C_{t-1}$. Note that at round $t=1$, $C_{t-1}=C_0=C$, i.e., the input set of clients (line \ref{lst:fl-line:client_initialisation}). 

\begin{algorithm}[h!]
	\caption{Adaptive federated learning process.}
	\label{lst:federated-training}
	\begin{algorithmic}[1]
		\renewcommand{\algorithmicrequire}{\textbf{Input:}}
		\renewcommand{\algorithmicensure}{\textbf{Output:}}
		\Require Global model ($w_0$), set of clients $(C)$
		\Procedure{AdaptiveFederatedTraining}{}
		\State $a_{max}\gets 0$ \Comment Max accuracy score
		\State $sc \gets 0$ \Comment Early stop counter
		\State $C_0\gets C$ \label{lst:fl-line:client_initialisation}
		\State $c_e=e_{max}, c_s=s_{max}\ \forall c\in C_0$ \Comment Epochs and steps \label{lst:fl-line:emin_emax}
		\State $c\gets\Call{InitClients}{w_0, c_e, c_s}\ \forall c\in C_0$
		\For {round $t=1,2,3,...$} \Comment Federated training loop
		\ForAll {$c \in C_{t-1}$} \Comment In parallel \label{lst:fl-line:update_start}
		\State $w_t^c \gets\Call{ClientUpdate}{w_{t-1}, c_e, c_s}$
		\EndFor \label{lst:fl-line:update_stop}
		\vspace{0.5mm}
		\State $w_t = \frac{1}{|C|}\sum_{c=1}^{|C|} w_t^c$  \Comment Arithmetic mean \label{lst:fl-line:arithmetic_mean}
		\vspace{1mm}
		\State $a^{\mu}\gets [a^c]_{c\in C} \gets$ \Call{SendModel}{$w_t$, $C$} \label{lst:fl-line:send_model}
		\vspace{0.5mm}
		\If {$a^{\mu} > a_{max}$}\label{lst:fl-line:evaluate_f1_start}
		\State  $\bar{w}\gets w_t$ \Comment Save best model
		\State $a_{max} \gets a^{\mu}$ \Comment Save max accuracy score
		\State $sc \gets 0$ \Comment Reset early stop counter
		\Else 
		\State $sc \gets sc+1$
		\EndIf		\label{lst:fl-line:evaluate_f1_stop}
		\If {$sc > \textsc{patience}$} 
		\State \Call{SendModel}{$\bar{w}$, $C$}  \Comment Send final model  \label{lst:fl-line:final_model}
		\State \textbf{return} \Comment End of the process
		\Else
		\State $C_{t} \gets \Call{SelectClients}{C, [a^c]_{c\in C}, a^{\mu}}$ \label{lst:fl-line:update_parameters}
		\EndIf
		\EndFor \label{lst:fl-line:final_loop_stop}
		\EndProcedure
	\end{algorithmic}
\end{algorithm}

At each round, the server computes the average of the parameters from all clients, regardless of whether they were involved in the previous round of training (line \ref{lst:fl-line:arithmetic_mean}). Please note that to speed up convergence on unbalanced and non-\ac{iid} data across clients, \ac{ourtool} replaces the weighted mean in Equation \ref{eq:fed_avg} with the arithmetic mean, similarly to other works in literature (e.g., \cite{stc,flddos,tian2021lightweight}). The new global model is sent to all clients, which return the accuracy scores  $[a^c]_{c\in C}$ obtained on their local validation sets with the new global model (line \ref{lst:fl-line:send_model}). The server computes the mean accuracy score value $a^{\mu}$, which is used to evaluate the progress of the federated training (lines \ref{lst:fl-line:evaluate_f1_start}-\ref{lst:fl-line:evaluate_f1_stop}). If $a^{\mu} > a_{max}$, the new global model is saved and the stopping counter $sc$ is set to $0$. Otherwise, $sc$ is increased by one to record no improvements. 
When $sc > \textsc{patience}$ (in our experiments we set  $\textsc{patience}=25$ rounds), the process stops and the best model is sent to all the clients for integration in their \acp{ids} (line \ref{lst:fl-line:final_model}). Otherwise, the server calls Algorithm \ref{lst:update_parameters} to determine which clients will participate in the next round and to assign the number of epochs and \ac{mbgd} steps to each of them.

\begin{algorithm}[h!]
	\caption{Select the clients for the next round of training.}
	\label{lst:update_parameters}
	\begin{algorithmic}[1]
		\renewcommand{\algorithmicrequire}{\textbf{Input:}}
		\renewcommand{\algorithmicensure}{\textbf{Output:}}
		\Require Clients ($C$), accuracy scores ($[a^c]_{c\in C}$), average accuracy score ($a^{\mu}$)
		\Ensure List of selected clients  $(C')$
		\Procedure{SelectClients}{$C$, $[a^c]_{c\in C}$, $a^{\mu}$}
		\State $C'\gets \{c \in C\mid a^c \le a^{\mu}\}$ \label{lst:up-line:clients_subset}
		\State $\underline{a} = min_{c\in C'}(a^c)$
		\State $\overline{a} = max_{c\in C'}(a^c)$
		\ForAll {$c \in C'$}\label{lst:up-line:parameters_loop_start}
		\vspace{1.5mm}
		\State $\textstyle \sigma = \frac{\overline{a}-a^c}{\overline{a}-\underline{a}}$ \Comment Scaling factor
		\vspace{0.5mm}
		\State $c_e=e_{min}+(e_{max}-e_{min})\cdot \sigma$
		\State $c_s=s_{min}+(s_{max}-s_{min})\cdot \sigma$
		\EndFor\label{lst:up-line:parameters_loop_stop}
		\State $\textbf{return}\ C'$
		\EndProcedure
	\end{algorithmic}
\end{algorithm}

Algorithm \ref{lst:update_parameters} starts with selecting the subset of clients $C'$ that will execute the local training in the next round. $C'$ is the set of $c\in C$ whose accuracy score $a^c$ obtained on their local validation set is lower than the mean value $a^{\mu}$ (line \ref{lst:up-line:clients_subset}). The number of epochs and steps assigned to each client $c\in C'$ depends on the value of $a^c$. The rationale is that the higher $a^c$, the lower the amount of computation needed from the client (thus, fewer epochs and \ac{mbgd} steps/epoch, as explained at the beginning of this section). 
This is formalised in the equations within the loop at lines \ref{lst:up-line:parameters_loop_start}-\ref{lst:up-line:parameters_loop_stop}, where each client $c\in C'$ is assigned a minimum number of epochs/steps plus an additional amount that is inversely proportional to the accuracy score $a^c$. The scale factor $\sigma$ ranges over $[0,1]$, assuming value $0$ when $a^c = max_{c\in C'}(a^c)$ (hence $c_e=e_{min}$ and $c_s=s_{min}$) and value $1$ when $a^c = min_{c\in C'}(a^c)$ (hence $c_e=e_{max}$ and $c_s=s_{max}$). Algorithm \ref{lst:update_parameters} returns the set of clients $C'$ that will perform computation during the next round, each assigned with a specific number of epochs and \ac{mbgd} steps.

\begin{algorithm}[h!]
	\caption{Local training procedure at client $c$.}
	\label{lst:client-update}
	\begin{algorithmic}[1]
		\renewcommand{\algorithmicrequire}{\textbf{Input:}}
		\renewcommand{\algorithmicensure}{\textbf{Output:}}
		\Require Global parameters $w$, epochs $(c_e)$, MBGD steps $(c_s)$
		\Ensure Updated parameters $(w)$
		\Procedure{ClientUpdate}{$w$, $c_e$, $c_s$}
		\State $X,y\gets\Call{LoadDataset}{ }$
		\If {$c_s > 0$} 
		\State $c_b\gets max(|X_{train}|/c_s,1)$ \Comment Compute batch size \label{lst:cu-line:batch_size}
		\EndIf
		\State $\mathcal{B}\gets$ split $X_{train}$ into batches of size $c_b$
		\For{epoch $e$ from 1 to $c_e$} \label{lst:cu-line:gs_loop_start}
		\ForAll{batch $b\in \mathcal{B}$}
		\State $w\gets w-\eta\nabla L(w,b)$
		\EndFor
		\EndFor  \label{lst:cu-line:gs_loop_stop}
		\State \textbf{return} $w$ \Comment Return updated parameters to server
		\EndProcedure
	\end{algorithmic}
\end{algorithm}

The pseudo-code provided in Algorithm \ref{lst:client-update} outlines the local training procedure carried out by clients. This process starts from the weights and biases of the current global model $w$ received from the server, and is executed for a number of epochs $c_e$ and \ac{mbgd} steps $c_s$ assigned by the server.
The first operation is the computation of the batch size $c_b$ using $c_s$ (line \ref{lst:cu-line:batch_size}). It ensures that $c_b\ge 1$, for the cases in which the number of samples in the local training set is smaller than $c_s$. Once the batch size is computed, the algorithm continues with $c_e\cdot c_b$ steps of gradient descend (lines \ref{lst:cu-line:gs_loop_start}-\ref{lst:cu-line:gs_loop_stop}) and finally returns the updated model to the server.

%% file: dataset.tex
\section{The Dataset}\label{sec:dataset}

\begin{table*}[bp]
	\caption{Overview of the CIC-DDoS2019 \ac{ddos} attack types.} 
	\label{tab:unb-dataset}
	\small 
	\centering 
	\begin{tabular}{p{1.6cm}p{1.1cm}p{1.4cm}p{12.3cm}} \toprule[\heavyrulewidth]
		\textbf{Attack} &\textbf{\#Flows} &\textbf{Transport}  & \textbf{Description} \\ \midrule[\heavyrulewidth]
		\textbf{DNS}  & 441931 & \multirow{6}{1.1cm}{UDP} & \multirow{6}{12.5cm}{\ac{ddos} attacks that exploit a specific UDP-based network service to overwhelm the victim with responses to queries sent by the attacks to a server using the spoofed victim's IP address. Six types of network services have been exploited to generate these attacks: \ac{dns}, \ac{ldap}, \ac{mssql}, \ac{ntp}, \ac{netbios}  and \ac{portmap}.} \\ 
		\textbf{LDAP}   & 11499 &&  \\ 
		\textbf{MSSQL}   & 9559537 && \\ 
		\textbf{NTP}   & 1194836 &&  \\ 
		\textbf{NetBIOS}   & 7553086 &&  \\ 
		\textbf{Portmap} & 186449 && \\  \midrule
		\textbf{SNMP}   & 1334534 &  UDP & Reflected amplification attack leveraging the \ac{snmp} protocol (UDP-based) used to configure network devices. \\  \midrule
		\textbf{SSDP}   & 2580154 & UDP & Attack based on the \ac{ssdp} protocol that enables UPnP devices to send and receive information over UDP. Vulnerable devices send UPnP replies to the spoofed IP address of the victim.\\  \midrule 
		\textbf{TFTP}  & 6503575 & UDP & Attack built by reflecting the files requested to a \ac{tftp} server toward the victim's spoofed IP address.  \\ \midrule
		\textbf{Syn Flood} & 6056402 & TCP & Attack that exploits the TCP three-handshake mechanism to consume the victim's resources with a flood of SYN packets. \\  \midrule
		\textbf{UDP Flood}  & 6969476 & UDP & Attack built with high rates of small spoofed UDP packets with the aim to consume the victim's network resources.  \\  \midrule
		\textbf{UDPLag}  & 474018 & UDP & UDP traffic generated to slow down the victim's connection to the online gaming server. \\  \midrule
		\textbf{WebDDoS}  & 146 & TCP & A short \ac{ddos} attack (around 3100 packets) against a web server on port 80. \\  \midrule
		\textbf{Total} & 42865789 && Despite the huge amount of flows, the dataset is heavily imbalanced, containing 8 predominant \ac{ddos} attack types, with more than one million flows each, a few tenths of thousands flows for the LDAP and Portmap reflection attacks, and only 146 flows for the WebDDoS attack. \\  	\bottomrule[\heavyrulewidth]
	\end{tabular}
\end{table*}

\ac{ourtool} is validated with a recent dataset of \ac{ddos} attacks, CIC-DDoS2019 \cite{sharafaldin2019developing}, provided by the Canadian Institute of Cybersecurity of the University of New Brunswick. CIC-DDoS2019 consists of several days of network activity, and includes both benign traffic and 13 different types of \ac{ddos} attacks. The dataset is publicly available in the form of pre-recorded traffic traces, including full packet payloads, plus supplementary text files containing labels and statistical details for each traffic flow \cite{cicddos2019}. The benign traffic of the dataset has been generated using the B-profile introduced in \cite{sharafaldin2018towards}, which defines distribution models for web (HTTP/S), remote shell (SSH), file transfer (FTP) and email (SMTP) applications. Instead, the attack traffic has been generated using third-party tools and can be broadly classified into two main categories: \textit{reflection-based} and \textit{exploitation-based} attacks. The first category includes those attacks, usually based on the UDP transport protocol, in which the attacker elicits responses from a remote server (e.g., a DNS resolver) towards the spoofed IP address of the victim. Hence, the victim is ultimately overwhelmed by the server's replies. The second category relates to those attacks that exploit known weaknesses of some network protocols (e.g., the three-way handshake of TCP). An overview of the  CIC-DDoS2019 dataset is provided in Table \ref{tab:unb-dataset}.

In Table \ref{tab:unb-dataset}, the column \textit{\#Flows} indicates the amount of bi-directional TCP sessions or UDP streams contained in the traffic traces provided with the dataset, each flow identified by a 5-tuple (source IP address, source TCP/UDP port, destination IP address, destination TCP/UDP port and IP protocol). Before experimenting with our solution, we have pre-processed the traffic traces with the tool developed in our previous work LUCID \cite{doriguzzi2020lucid}. The resulting representations of traffic flows are in the form of arrays of shape $n=10$ rows and $f=11$ columns. Each row contains a representation of a packet based on 11 features, the same considered in the LUCID paper: \textit{Time}, \textit{Packet Length}, \textit{Highest Protocol}, \textit{IP Flags}, \textit{Protocols}, \textit{TCP Length}, \textit{TCP Ack}, \textit{TCP Flags}, \textit{TCP Window Size}, \textit{UDP Length} and \textit{ICMP Type}. If the number of packets of a flow is lower than $n$, the array is zero-padded. The number of non-zero rows in the array can be seen as another feature that we call \textit{FlowLength}.
It is worth recalling that packets are inserted into the array in chronological order and that the timestamp is the inter-arrival time between a packet and the first packet in the array. As the packet attributes are extracted using TShark \cite{TShark}, we can use some high-level features such as the highest protocol detected in the packet and the list of all protocols recognised in the packet.
The LUCID dataset parser splits each traffic flow into smaller subsets of packets to produce samples that are consistent with real-world settings, where the detection algorithms must cope with fragments of ﬂows collected over pre-defined time windows. 
Shorter time windows allow faster decisions, but also a higher fragmentation of the flows, hence a possible decrease in the classification accuracy. 
In this work, we use a time window of duration 10 seconds for both benign and attack traffic. By taking this choice, we slightly increase the size of the smallest attack (202 WebDDoS samples obtained by splitting 146 flows), while maintaining an adequate level of accuracy, as per evaluation results reported in the LUCID paper.

\begin{figure}[t!]
	\begin{center}
		\includegraphics[width=0.95\linewidth]{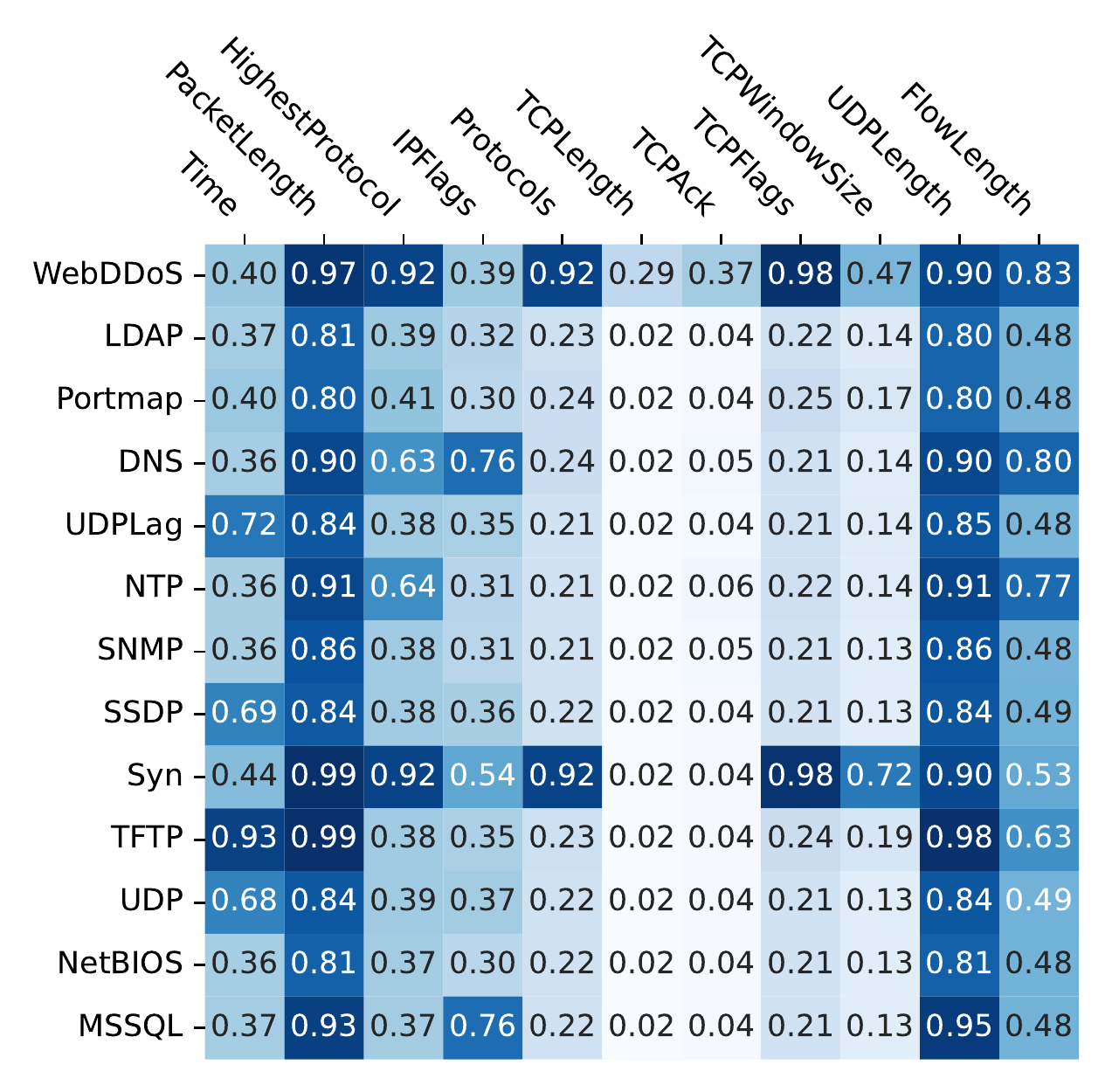}
		\caption{\ac{jsd} map of the probability distributions of the features.}
		\label{fig:distance_matrix}
	\end{center}
\end{figure}

\subsection{Feature distribution analysis}\label{sec:feature_distribution}
In the use-case scenario considered in this work, individual clients contribute to the federated training with private data collections of benign and attack traffic, possibly drawn from non-identical feature distributions. To compare the feature distributions among the attack types of the CIC-DDoS2019 dataset, we use the \ac{jsd} \cite{jsd} metric. \ac{jsd} measures the degree of overlapping of two probability distributions, where distance zero means identical distributions, while distance one means that the two distributions are supported on non-overlapping domains.  
Figure \ref{fig:distance_matrix} reports the \ac{jsd} values for all features and attack types. More precisely, an element $(i,j)$ in the matrix is the average \ac{jsd} value between the attack at row $i$ and each of the other attacks, computed on their probability distributions of the feature at column $j$.

\begin{figure}[t!]
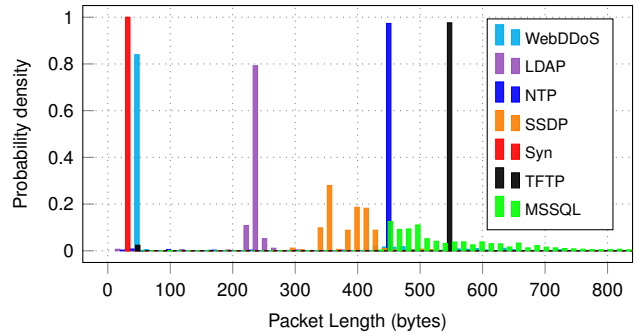

	\chartPacketLengthDistribution
	\caption{Probability density functions of the \textit{Packet Length} feature.}
	\label{fig:packet_length}
\end{figure}

In Figure \ref{fig:distance_matrix}, we observe that every attack presents at least one feature whose probability distribution domain is almost disjoint from those of the other attacks. As also shown in Figure \ref{fig:packet_length}, this primarily relates to features \textit{Packet Length} and \textit{UDP Length} (redundant in this dataset, where most of the attacks are UDP-based). Indeed, similar distributions with different packets sizes can be observed on \textit{LDAP}, \textit{NTP} and \textit{TFTP} attacks, while the packet sizes in other attacks are distributed across larger domains (Figure \ref{fig:packet_length}). About TCP-based attacks, all the \textit{Syn Flood} packets have \textit{Packet Length} equal to 40 bytes, while more than 80\% of the \textit{WebDDoS} packets have a size of either 66 or 74 bytes.

Similar considerations apply to other features such as \textit{Flow Length}, indicating that also the distribution of packets/sample changes across the different attacks. Finally, it is worth noting the large \ac{jsd} distance of the two TCP-based attacks (WebDDoS and Syn Flood) from the other attacks (UDP-based) on most of the features. We will show in Section \ref{sec:evaluation} the negative impact of such out-of-distribution attacks on the convergence of \ac{fvg}.

%% file: setup.tex
\section{Experimental Setup}\label{sec:setup}
The \ac{ourtool} approach is validated using a fully connected neural network model (or \ac{mlp}), which is initialised with random parameters (weights and biases) by the server and locally trained multiple times by the clients, as per the procedure presented in Section \ref{sec:methodology}. 

As we are interested in measuring the benefits of \ac{ourtool} over other approaches (\ac{fvg} and FLDDoS \cite{flddos}) in terms of load on the clients, convergence time and classification accuracy, we want to avoid the impact of communication inefficiencies, such as network latencies that can occur in distributed deployments. Therefore, \ac{ourtool} is implemented as a single Python process using Tensorflow 2.7.1 \cite{tensorflow}, thus server and clients are executed on the same machine and communicate through local procedure calls. Please note that this implementation choice does not affect the validity or generality of our work. Federated training and model testing have been performed on a server-class computer equipped with two 16-core Intel Xeon Silver 4110 @2.1 GHz CPUs and 64 GB of RAM.

\subsection{Dataset preparation}\label{sec:dataset_preparation}
The CIC-DDoS2019 dataset has been split into $13$ smaller datasets, each containing samples of benign traffic and only one type of attack. Furthermore, we deliberately introduced an imbalance across the $13$ datasets by doubling the number of samples from one dataset to another, starting from the one with the smallest attack (202 WebDDoS samples) and culminating in the largest dataset (MSSQL), which has been reduced to 819204 attack samples. Every dataset split has been carefully balanced to ensure an approximately equal distribution between benign and \ac{ddos} samples. The balanced datasets were further divided into training (90\%) and test (10\%) sets, with an additional 10\% of the training set reserved for validation purposes. (Table \ref{tab:datasets-splits}).
\begin{table}[h!]
	\caption{CIC-DDoS2019 dataset splits.} 
	\label{tab:datasets-splits}
	\small 
	\centering 
	\begin{threeparttable}
		\begin{tabular}{lcccc} \toprule[\heavyrulewidth]
			\begin{tabular}{@{}l@{}}\textbf{Dataset split}\end{tabular} & \begin{tabular}{@{}c@{}}\textbf{Samples}\end{tabular} & \begin{tabular}{@{}c@{}}\textbf{Training}\end{tabular} & \begin{tabular}{@{}c@{}}\textbf{Validation}\end{tabular}& \begin{tabular}{@{}c@{}}\textbf{Test}\end{tabular} \\ \midrule[\heavyrulewidth]
			
			\textbf{WebDDoS} & 402 & 321 & 37 & 44\\   
			\textbf{LDAP} & 854 & 633 & 135 & 86\\ 
			\textbf{Portmap} & 1605 & 1299 & 145 & 161\\ 
			
			\textbf{DNS} & 3207 & 2595 & 291 & 321 \\   
			\textbf{UDPLag} & 6400 & 5184 & 576 & 640\\ 
			\textbf{NTP} & 12807 & 10372 & 1153 & 1282\\ 
			
			\textbf{SNMP} & 25649 & 20775 & 2309 & 2565\\   
			\textbf{SSDP} & 51207 & 41477 & 4609 & 5121\\ 
			\textbf{Syn Flood} & 102400 & 82940 & 9216 & 10244\\
			\textbf{TFTP} & 204800 & 165887 & 18433 & 20480\\ 
			\textbf{UDP Flood} & 409601 & 331772 & 36864 & 40965\\
			\textbf{NetBIOS} & 819200 & 663551 & 73728 & 81921\\ 
			\textbf{MSSQL} & 1638404 & 1327105 & 147457 & 163842\\
			\bottomrule[\heavyrulewidth]
		\end{tabular}
	\end{threeparttable}
\end{table}

The dataset splits outlined in Table \ref{tab:datasets-splits} serve as an evaluation framework for \ac{ourtool} under a worst-case scenario. In this scenario, each attack is exclusively assigned to a single client (one-to-one mapping), resulting in a pathological non-\ac{iid} partition of the data, as referred to by McMahan et al. \cite{mcmahan2017communication}. Furthermore, these splits can be combined to create larger federations to assess scalability or to replicate experimental settings employed by other state-of-the-art approaches for comparison purposes.

\subsection{\ac{ann} architecture}
\begin{figure}[h!]
	\begin{center}
		\includegraphics[width=1\linewidth]{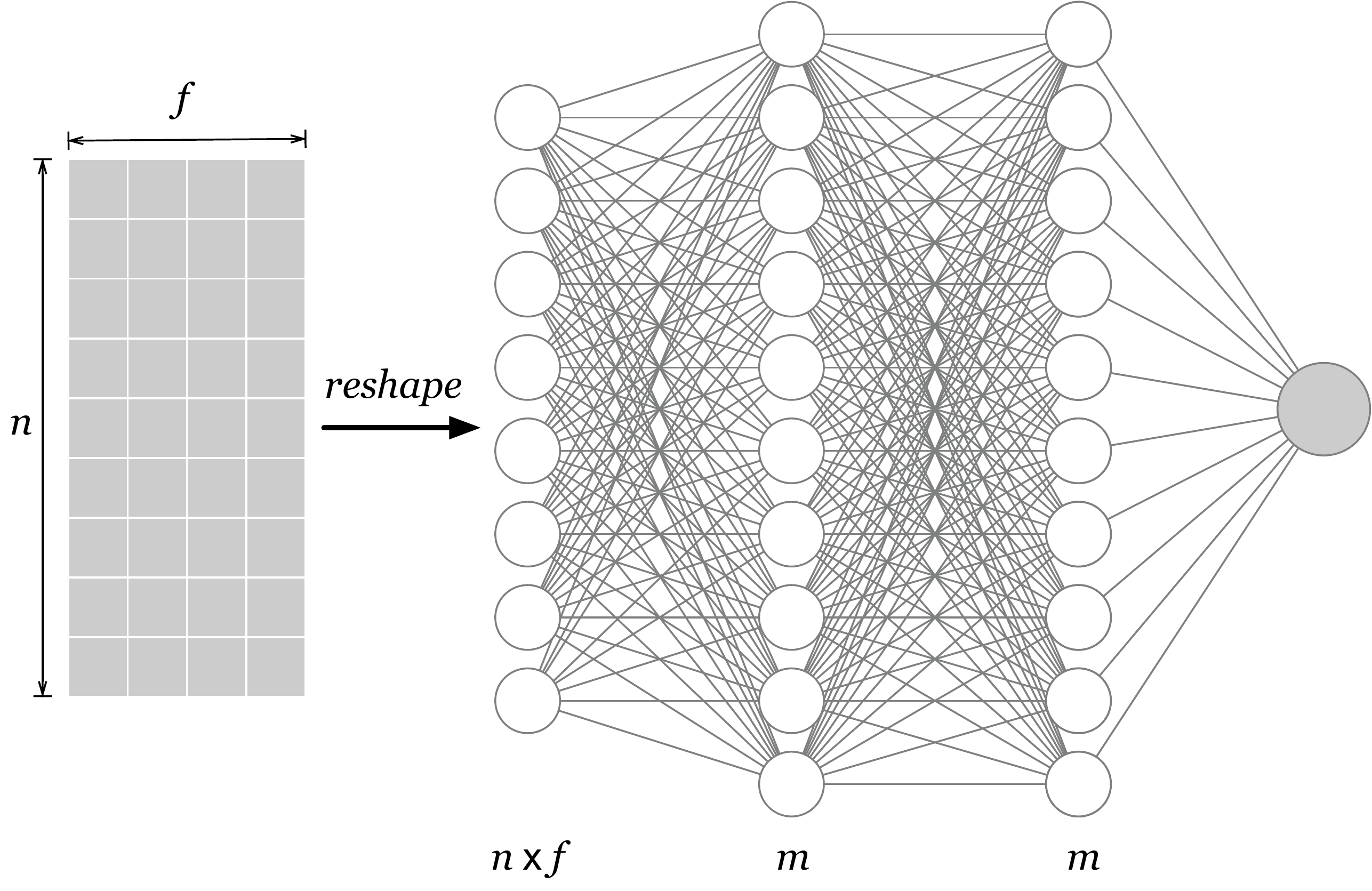}
		\caption{Architecture of the \ac{ann} used to evaluate \ac{ourtool}.}
		\label{fig:architecture}
	\end{center}
\end{figure}

The architecture of our \ac{mlp} model consists of an input layer of shape $n\times f$ neurons, a single-neuron output layer and $l$ hidden dense layers of $m$ neurons each (Figure \ref{fig:architecture}). The input of the neural network is an array-like representation of a traffic flow, where lines are packets of the flow in chronological order from top to bottom, and columns are packet-level attributes (see Section \ref{sec:dataset}). Before processing, each array is re-shaped into a $n\cdot f$-size vector, where packets are lined up one after another in chronological order.  
The objective of the local training procedure, summarised in Algorithm \ref{lst:client-update}, is to minimise the cross-binary cost function defined in Equation \ref{eq:cross-entropy}. The cost function measures the quality of the model's traffic classification compared to the ground truth of the input. At each training epoch, the error is back-propagated through the network and it is used to iteratively update the model's weights until convergence is obtained.  
\begin{equation}\label{eq:cross-entropy}
	c = -\frac{1}{s}\sum_{j=1}^{s}(y_j\log p_j + (1-y_j)\log(1-p_j)) 
\end{equation}
In Equation \ref{eq:cross-entropy}, $y_j$ is the ground truth label of each input flow $j$ in a batch of $s$ samples. The label of benign flows is equal to $0$, while the label of \ac{ddos} flows is equal to $1$. The value of $p_j\in (0,1)$ is the predicted probability flow $j$ is \ac{ddos}. The cost $c$, as computed in Equation \ref{eq:cross-entropy}, tends to $0$ when the output probabilities of the flows are close to the respective ground truth labels.    

Note that, the cost is computed by each client independently, using only local training data. In the case of non-\ac{iid} features across clients (cf. Section \ref{sec:feature_distribution}), the distance between the data distributions can lead the weights of different clients to diverge, slowing down the convergence of the \ac{fl} process to the target performance of the global model \cite{zhao2018federated}.

\subsection{\ac{fl} hyper-parameters}\label{sec:hyperparameters}
The whole \ac{fl} process is configured with a set of hyper-parameters, which have been determined either based on the results of the preliminary tuning activities (\textsc{patience}, \ac{mlp} architecture), or based on the observations of McMahan et al. \cite{mcmahan2017communication} on local training epochs and batch size.
The hyper-parameters presented in Table \ref{tab:hyper-parameters} have been used to validate \ac{ourtool} and to compare it against \ac{fvg} and \ac{flddos} \cite{flddos}, a recent \ac{fvg}-based solution for \ac{ddos} attack detection introduced in Section \ref{sec:related} and presented below. The values in the table have been kept constant across all experiments described in Section \ref{sec:evaluation}. 

\begin{table}[h!]
	\caption{Hyper-parameters of \ac{ourtool}.} 
	\label{tab:hyper-parameters}
	\small 
	\centering 
	\begin{adjustbox}{width=1\linewidth}
	\begin{threeparttable}
		\begin{tabular}{lcp{5.2cm}} \toprule[\heavyrulewidth]
			\textbf{Name} & \textbf{Value} & \textbf{Description} \\ \midrule[\heavyrulewidth]
			\textbf{\textsc{patience}} & 25 & Max \ac{fl} rounds with no progress. \\  
			\textbf{Min epochs} & 1 & Min number of local training epochs. \\ 
			\textbf{Max epochs} & 5 & Max number of local training epochs. \\ 
			\textbf{Min steps} & 10 &  Min number \ac{mbgd} steps. \\ 
			\textbf{Max steps} & 1000 & Max number \ac{mbgd} steps. \\ 
			\textbf{$n\times f$} & $10\times 11$ & Size of the \ac{mlp} input layer. \\
			\textbf{$l$} & 2 & Number of hidden layers. \\
			\textbf{$m$} & 32 & Number of neurons/layer. \\
			\bottomrule[\heavyrulewidth]
		\end{tabular}
	\end{threeparttable}
	\end{adjustbox}
\end{table}

The value of \textsc{patience} has been set to 25 rounds, which, compared to lower values, guarantees good accuracy on small and non-\ac{iid} attacks, such as WebDDoS and Syn Flood. In terms of the number of hidden layers and activations, we started with larger architectures and then progressively reduced the dimensions until we reached a configuration that allowed good detection accuracy on all attacks in a reasonable time.
The dynamic tuning of epochs and \ac{mbgd} steps implemented by \ac{ourtool} is configured with the ranges presented in Table \ref{tab:hyper-parameters}. The minimum and maximum values of epochs have been set as the same values used to evaluate \ac{fvg}. Unlike other approaches such as \ac{fvg} and \ac{flddos}, we do not specify the batch size, but instead, we tune the number of \ac{mbgd} steps each client has to take at each round, hence controlling the client's training process without having to consider the size of its local dataset. We experiment with amounts of steps ranging between $10$ and $1000$.

Concerning our implementation of \ac{fvg} and \ac{flddos}, we use the same \ac{mlp} architecture used for testing \ac{ourtool}, with the same values of hyper-parameters $n$, $f$, $l$ and $m$ reported in Table \ref{tab:hyper-parameters}. The remaining hyperparameters, such as the number of local epochs $E$ and batch size $B$ (described in Section \ref{sec:problem_formulation}), are set according to the values specified in the respective papers \cite{mcmahan2017communication, flddos}.

In the case of \ac{fvg}, we experiment with $E=1$ and $E=5$ epochs/round of local training and with a fixed batch size of $B=50$ samples. Other values could be also chosen (e.g., $E=20$ or $B=10$, also used by McMahan et al. in their work), but after a preliminary investigation, we found that the little gain in accuracy achieved with further \ac{mbgd} steps was not sufficient to balance the impressive amount of additional local computation on clients with large datasets. 
In their study, McMahan et al. assess the performance of \ac{fvg} using a client fraction $F=0.1$. For their evaluation, they form federations consisting of 100, 600, and 1000+ clients, focusing on tasks such as image classification, digit recognition, and language modelling. However, in our experiments, we primarily work with smaller federations ranging from 13 to 90 clients. Therefore, we align with \ac{flddos} authors' recommendation for testing \ac{fl} in a DDoS attack detection scenario and we set $F$ to 0.8.	

\ac{flddos} aims to mitigate the limitations of \ac{fvg} on non-\ac{iid} \ac{ddos} attack data by maintaining a local model at each client. 
More precisely, at round $t$, each client $c\in C$ updates the global model $w_t^c$ as done with \ac{fvg}. In addition, the client maintains a local model $v_t^c$ that is trained solely with the local data. These two models are then merged to create a \textit{personalised model} $\bar{v}_t^c$, which is subsequently sent to the server for aggregation. The formula used to calculate the personalised model is as follows:
\begin{equation}
	\bar{v}_t^c = \gamma^c w_t^c + (1- \gamma^c)v_t^c\quad\forall c\in C
\end{equation}

Where the weighting factors $\gamma^c$ can be tuned based on the level of heterogeneity in the clients' data. Note that when $\gamma^c=1$ $\forall c\in C$, FLDDoS corresponds to \ac{fvg}. Otherwise, by setting $0\le \gamma^c<1$ for $c\in C$, one can allow a client to tune the relative importance of the local model $v_t^c$ in the personalised model that is sent to the server for aggregation.
Although the authors of \ac{flddos} do not suggest any possible values for $\gamma^c$, in our experiments we try to  
understand the impact of model personalization on the convergence of the \ac{fl} process. To this purpose, we test FLDDoS with $\gamma^c=1$ (no local model as for \ac{fvg}) for those clients contributing to the \ac{fl} with UDP-based attacks, and $\gamma^c=0.9$ for the clients with TCP-based attacks, namely WebDDoS and SYN Flood. As demonstrated in Section \ref{sec:convergence_analysis}, learning such attacks can be particularly challenging using the \ac{fvg} since these are the only two TCP-based attacks in the dataset, while the majority of attacks are UDP-based. By setting $\gamma^c=0.9$, we try to increase the contribution of the TCP-based attacks when building the global model and to understand whether the FLDDoS approach addresses the limitations of \ac{fvg}.

Additionally, we use the hyper-parameter values reported in the paper, including a fraction of clients $F=0.8$, a batch size of $B=100$ samples, and $E=10$ local epochs per round.

%% file: evaluation.tex
\section{Experimental Evaluation}\label{sec:evaluation}
The evaluation focuses on assessing the adaptive approach of \ac{ourtool} in terms of convergence time of the \ac{fl} process, \ac{ddos} attack detection accuracy of the global model, and scalability. For this purpose, we utilize the CIC-DDoS2019 dataset, which is configured according to the specifications outlined in Section \ref{sec:dataset_preparation}.

The classification performance of \ac{ourtool} is measured in terms of \textit{F1 score} and \textit{\ac{tpr}}. The \ac{tpr}, also called \textit{Recall}, is the ratio between the correctly detected \ac{ddos} samples and all the \ac{ddos} samples in the dataset. The \ac{tpr} quantifies how well a model can identify the \ac{ddos} attacks. 

The \textit{F1 Score} is a widely used metric to evaluate classifier accuracy, computed as the harmonic mean of \textit{Precision} and \ac{tpr}, with \textit{Precision} (\textit{Pr}) being the ratio between the correctly detected \ac{ddos} samples and all the detected \ac{ddos} samples. The \textit{F1 Score} is formally defined as $F1=2\cdot\frac{Pr\cdot TPR}{Pr + TPR}$. The \textit{F1 Score} is also used as the accuracy metric in the implementation of Algorithms \ref{lst:federated-training} and \ref{lst:update_parameters} presented in Section \ref{sec:methodology}. 

\subsection{State-of-the-art comparison}\label{sec:sota_comparison}
We compare \ac{ourtool} against \ac{fvg}, the original \ac{fl} algorithm proposed by McMahan et al. \cite{mcmahan2017communication} and against a recent \ac{fl}-based solution for \ac{ddos} attack detection called FLDDoS \cite{flddos}. Both \ac{fvg} and \ac{flddos} adopt a randomised client selection strategy, while also employing fixed batch sizes and local training epochs across all clients.
The goal of this evaluation is to expose the limitations of such design choices in a cybersecurity scenario, where the server does not possess a test set (for the reasons discussed earlier in this paper) to measure the performance of the global model on different attack types.  

\subsubsection{Convergence analysis}\label{sec:convergence_analysis}
In this experiment we train the global model with \ac{ourtool} until convergence, i.e., waiting for \textsc{patience}=25 rounds with no progress in the average F1 Score across the clients. Following this, we evaluate the performance of the original \ac{fvg} algorithm and FLDDoS by subjecting them to the same number of training rounds as \ac{ourtool}. 

We perform the convergence analysis in a worst-case scenario, i.e., with a federation of $13$ clients and a one-to-one mapping between clients and \ac{ddos} attack types. We replicate the same experiment by employing a federation of $50$ clients, each containing two attack types in their local dataset. The latter settings align with Lv et al.'s evaluation of FLDDoS in their study \cite{flddos}.

Each experiment is repeated $10$ times and the average metrics are reported in this section. As TensorFlow relies on a pseudo-random number generator to initialise the global model, and both \ac{fvg} and FLDDoS perform a random selection of clients at each \ac{fl} round, each experiment is initiated with a unique random seed to ensure diverse testing conditions.

\begin{table}[h!]
	\caption{Average metrics over $10$ experiments in the $13$-client scenario, with one-to-one clients/attacks mapping.} 
	\label{tab:worst_case_comparison}
	\centering 
	\renewcommand{\arraystretch}{1.1}
	\begin{adjustbox}{width=0.49\textwidth}
	\begin{tabular}{>{\bfseries}l cccc} \toprule[\heavyrulewidth]
		\textbf{Metric} & \begin{tabular}{@{}c@{}}\textbf{\ac{ourtool}}\\ \textbf{E=A,S=A}\end{tabular}  & \begin{tabular}{@{}c@{}}\textbf{FedAvg}\\ \textbf{E=1,B=50}\end{tabular} & \begin{tabular}{@{}c@{}}\textbf{FedAvg}\\ \textbf{E=5,B=50}\end{tabular} & \begin{tabular}{@{}c@{}}\textbf{\ac{flddos}}\\ \textbf{E=10,B=100}\end{tabular}\\ \midrule[\heavyrulewidth]
		FL Rounds & 68 & 68 & 68 & 68 \\
		Round Time (sec) & 9.08 & 34.19 & 179.48 & 205.39 \\
		Total Time (sec) & 617 & 2325 & 12205 & 13967 \\
		F1 Score & 0.9667 & 0.8577 & 0.9157 & 0.9091 \\
		F1 StdDev & 0.0369 & 0.2714 & 0.1597 & 0.1605 \\
		F1 WebDDoS & 0.8990 & 0.0815 & 0.8148 & 0.7376 \\
		F1 Syn & 0.9877 & 0.4563 & 0.4613 & 0.5094 \\
		\bottomrule[\heavyrulewidth]
	\end{tabular}
	\end{adjustbox}	
\end{table}

\begin{figure*}[h!]
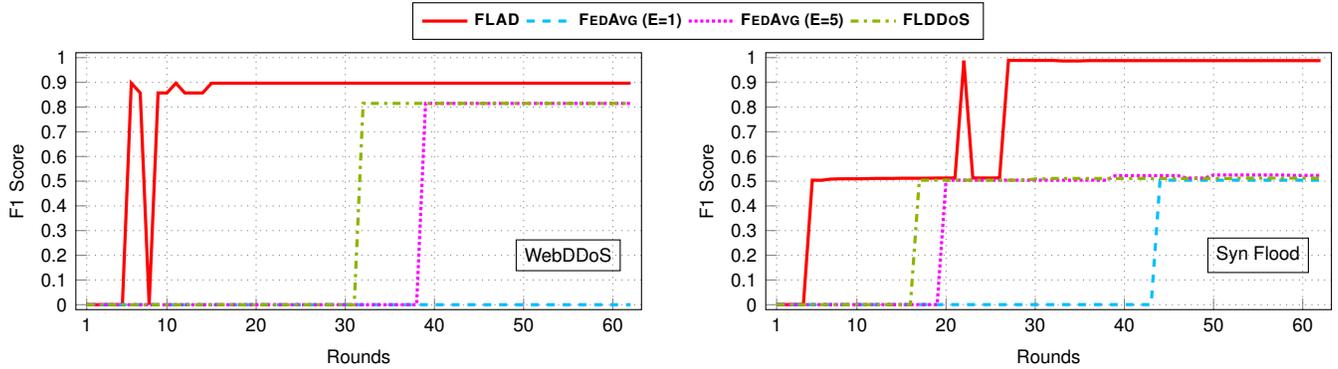

	\begin{subfigure}[t]{0.5\textwidth} 
		\chartFScoreWebDDoS
	\end{subfigure}
	\begin{subfigure}[t]{0.5\textwidth} 
		\chartFScoreSyn
	\end{subfigure}
	\caption{Performance on \acl{ood} data, namely on the WebDDoS and Syn Flood attacks.}
	\label{fig:ood_comparison}
\end{figure*}

The results obtained in the worst-case scenario are summarised in Table \ref{tab:worst_case_comparison}, which reports average metrics across the $10$ iterations of this experiment. As introduced in Section \ref{sec:hyperparameters}, \ac{ourtool} is configured with adaptive tuning of epochs and \ac{mbgd} steps of local training (E=A,S=A). \ac{ourtool} is compared against two configurations of \ac{fvg}, with E=1 and E=5 epochs/round of local training, and against \ac{flddos} configured with E=10. 

The table shows the advantages of \ac{ourtool} over \ac{fvg} and \ac{flddos}: higher accuracy within a shorter time frame. These improvements can be attributed to the dynamic client selection strategy implemented by \ac{ourtool}. At each round of the federated training process, clients are chosen based on the performance of the current aggregated model on their local datasets. Consequently, \ac{ourtool} prioritizes attacks that are more challenging to learn, specifically the \ac{ood} attacks WebDDoS and Syn Flood. The clients with these attacks are selected more frequently for local training, with an average of approximately $44$ and $46$ rounds respectively out of a total of $68$ rounds, compared to an average of around 18 rounds for the clients with the other attacks.

In contrast, both \ac{fvg} and \ac{flddos} rely on random client selection, where each client is involved in approximately $77\%$ of the training rounds (around 52 rounds on average out of a total of 68 rounds), considering the client fraction $F=0.8$ used in our experiments.
This results in longer rounds due to the frequent inclusion of clients with large local datasets, even when their contribution is not essential for improving the accuracy of the aggregated model. 
Furthermore, \ac{ourtool} dynamically tunes the amount of computation assigned to the selected clients at each round of training, resulting in a significant reduction in the average local training time per round. Comparatively, \ac{ourtool} achieves an average local training time of around $9$ seconds per round, while the two configurations of \ac{fvg} require $34$ and $179$ seconds per round, respectively. The \ac{flddos} configuration, on the other hand, takes more than $200$ seconds per round. Consequently, \ac{ourtool}'s adaptive allocation strategy not only decreases the per-round training time but also effectively reduces the overall duration of the federated training process.

It is also worth noting that the overall performance of \ac{flddos} and \ac{fvg} with E=5 is similar, as they assign approximately the same amount of computation to the clients. Specifically, \ac{flddos} is configured with E=10 epochs local training (as in the original paper by Lv et al. \cite{flddos}), while \ac{fvg} uses E=5 epochs with twice the number of \ac{mbgd} steps/epochs due to the smaller batch size.
Additionally, the strategy employed by \ac{flddos} to handle non-\ac{iid} data does not yield significant improvements compared to \ac{fvg} in our evaluation scenario. 
In fact, the local models maintained by clients with \ac{ood} data, such as WebDDoS and Syn Flood attack traffic, do not contribute to improving the accuracy of the global model on such attacks but, instead, increase the total training time.
     
This is clearly shown in Figure \ref{fig:ood_comparison}, which shows the performance trend of \ac{ourtool}, \ac{fvg} and \ac{flddos} on the WebDDoS and Syn Flood attack traffic during the first of the $10$ iterations of the experiment. 
The two plots clearly demonstrate that \ac{ourtool} achieves faster learning and higher accuracy for both of these attacks, while \ac{flddos} and \ac{fvg} with E=5 exhibit similar trends.

In these plots, we can also observe the adaptive mechanism of \ac{ourtool} in action. Once the global model has learnt a client's data profile, \ac{ourtool} excludes such client from the next round of federated training. In the case of clients with \ac{ood} data, such as WebDDoS and Syn attack data, this behaviour might cause the model to forget what it has previously learnt on such attacks, as can be seen on both plots in the figure. However, this prompts \ac{ourtool} to reintegrate such clients in the subsequent rounds of the training process, ultimately leading to global convergence.

\begin{figure}[h!]
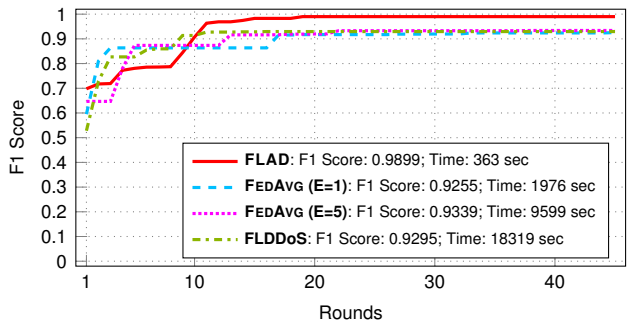

	\chartFScoreFiftyClients
	\caption{Performance comparison on a federation of $50$ clients, two attacks/client.}
	\label{fig:50_clients_comparison}
\end{figure}

Finally, Figure \ref{fig:50_clients_comparison} presents the performance trend of \ac{ourtool}, \ac{fvg}, and \ac{flddos} in a scenario with $50$ clients, where each client's dataset consists of two attacks along with benign traffic. Also in this case, we repeated the experiment $10$ times. However, due to limited space, only the test results of the first iteration are displayed. Nevertheless, a similar pattern was observed throughout the remaining nine iterations. The local datasets of the $50$ clients are generated by randomly combining pairs of the $13$ datasets listed in Table \ref{tab:datasets-splits}. For each test iteration, a different random seed is utilised to generate a distinct federation.

It is worth mentioning that, given these test settings, the two \ac{ood} attacks are present in the datasets of multiple clients. Due to this, we observe a higher average F1 score on the clients' validation sets compared to the $13$-client scenario (approximately $0.99$ with \ac{ourtool} and $0.94$ with \ac{flddos} and the two configurations of \ac{fvg}) and lower standard deviation across the $50$ local validation sets (approximately $0.01$ with \ac{ourtool} and $0.1$ with \ac{flddos} and \ac{fvg}). These results demonstrate the advantages of the adaptive mechanism implemented by \ac{ourtool}, even in scenarios with more uniformly distributed and less imbalanced data.

\subsubsection{Evaluation on unseen data}
To assess the performance of the global models trained using \ac{ourtool}, \ac{fvg}, and \ac{flddos}, we conduct evaluations on previously unseen data. To this purpose, we use the models trained in the worst-case scenario of $13$ clients (Section \ref{sec:convergence_analysis}). Thus, we test the capability of the global models to correctly recognise the 13 attacks. To this aim, Table \ref{tab:test_set_comparison_tpr} reports the average \ac{tpr} measured on the test sets of the clients using the final models obtained in the 10 experiments.
While \ac{ourtool} produces high \ac{tpr} values across all the attacks, the results of this experiment highlight the shortcomings of \ac{fvg} and \ac{flddos} when dealing with \acl{ood} attack traffic (the TCP-based attacks WebDDoS and Syn Flood). These findings validate the conclusions drawn from the aggregated metrics analysis presented in the previous section.

\begin{table}[h!]
	\caption{Comparison on clients' test sets (\ac{tpr}).} 
	\label{tab:test_set_comparison_tpr}
	\small 
	\centering 
	\begin{adjustbox}{width=0.49\textwidth}
	\begin{tabular}{>{\bfseries}l ccccc} \toprule[\heavyrulewidth]
		\textbf{Attack} & \begin{tabular}{@{}c@{}}\textbf{\ac{ourtool}}\\ \textbf{E=A,S=A}\end{tabular}  & \begin{tabular}{@{}c@{}}\textbf{FedAvg}\\ \textbf{E=1,B=50}\end{tabular} & \begin{tabular}{@{}c@{}}\textbf{FedAvg}\\ \textbf{E=5,B=50}\end{tabular} & \begin{tabular}{@{}c@{}}\textbf{\ac{flddos}}\\ \textbf{E=10,B=100}\end{tabular}\\ \midrule[\heavyrulewidth] \cline{3-5}
		WebDDoS & 0.7864 &  \multicolumn{1}{|c}{\textbf{0.0727}} & \textbf{0.7182} & \multicolumn{1}{c|}{\textbf{0.6455}} \\ \cline{3-5}
		LDAP & 0.9306 & 0.8972 & 0.9306 & 0.9083 \\
		Portmap & 0.9250 & 0.8548 & 0.9221 & 0.8740 \\
		DNS & 0.9779 & 0.9060 & 0.8799 & 0.8772 \\
		UDPLag & 0.9652 & 0.9978 & 0.9984 & 0.9981 \\
		NTP & 0.9660 & 0.9874 & 0.9701 & 0.9661 \\
		SNMP & 0.9574 & 0.9211 & 0.9586 & 0.9320 \\
		SSDP & 0.9663 & 0.9983 & 0.9988 & 0.9989 \\ \cline{3-5}
		Syn & 0.9767 & \multicolumn{1}{|c}{\textbf{0.3188}} & \textbf{0.3254} & \multicolumn{1}{c|}{\textbf{0.3861}} \\ \cline{3-5}
		TFTP & 0.9439 & 0.9483 & 0.9372 & 0.9524 \\
		UDP & 0.9656 & 0.9996 & 0.9995 & 0.9995 \\
		NetBIOS & 0.9218 & 0.8581 & 0.9272 & 0.8820 \\
		MSSQL & 0.9981 & 0.9994 & 0.9176 & 0.9503 \\
		\midrule
		Average & 0.9699 & 0.9396 & 0.9155 & 0.9232 \\
		\bottomrule[\heavyrulewidth]
	\end{tabular}
\end{adjustbox}
\end{table}

\subsubsection{Discussion}\label{sec:evaluation_discussion}
The key improvement of \ac{ourtool} over to \ac{fvg} (and other solutions based on it, such as \ac{flddos}) is the mechanism that monitors the performance of the global model, which allows the implementation of adaptive methods to control the \ac{fl} process and the definition of a stopping strategy. About the latter, in our experiments, we stop the \ac{fl} process after  \textsc{patience}=25 rounds with no progress in the average F1 score. Alternatively, more advanced stopping strategies are also possible with \ac{ourtool}. For instance, the practitioner might want to wait longer until a target average accuracy is reached, perhaps also combined with a target standard deviation of the accuracy scores to ensure that the performance is stable across all local datasets.

\subsection{Federated re-training with new attack data}\label{sec:retraining_results}
We now evaluate \ac{ourtool} in a realistic scenario, where the global model needs frequent retraining to learn new attacks.  

This experiment starts with two clients training on attack data (one attack type each) and benign traffic (Algorithm \ref{lst:federated-training} where $w_0$ is a set of randomly initialised parameters and $|C|=2$).  Once the federated training process converges, the resulting aggregated model is used as a starting point for another round of federated training (Algorithm \ref{lst:federated-training} with $w_0=\bar{w}$), in which we provide attack data (a new attack type) to a third client ($|C|=3$) to simulate the discovery of a new zero-day \ac{ddos} attack.  Once convergence is achieved, we restart training by introducing new attack data on a fourth client. This is repeated until all thirteen attacks have been added, one on each client, and all the clients are provided with a model that has been trained with all the attack profiles.  Each retraining iteration should converge as soon as possible and should produce aggregated models with high classification scores across the available attacks (high average F1 score with low standard deviation). 

At each step of this experiment, we stop the \ac{fl} process after \textsc{patience}=25 rounds with no progress in the average F1 score, and we start again by introducing a new attack as described above. 
As results may depend on the order in which attacks are introduced into the experiment, we repeat the whole experiment $10$ times, each time with a different sequence of attacks. 

\begin{figure}[!h]
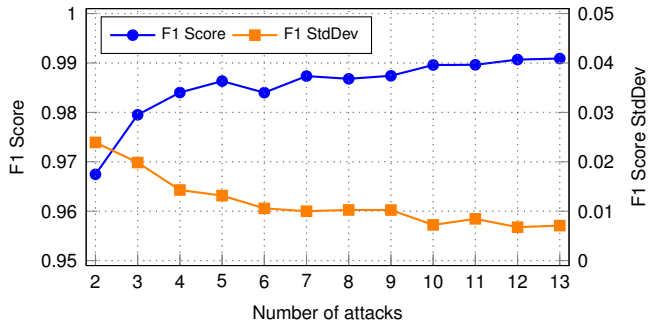

	\chartFScoreProgressive
	\caption{Mean and standard deviation of the F1 score at an increasing number of \ac{ddos} attacks.}
	\label{fig:f1_score_stddev_progressive}
\end{figure}

The experiment's findings have been visually presented in Figure \ref{fig:f1_score_stddev_progressive}.
The plot in the Figure captures the progression of the average F1 score and its corresponding standard deviation throughout the various stages of the experiment. The \textit{F1 Score} is the average F1 Score measured on the validations sets reported by the clients each time convergence is achieved at each step of the experiment (hence with 2, 3, \dots, 13 attacks) averaged over the 10 experiments). Similarly to the F1 score, we compute the \textit{F1 StdDev} as the average standard deviation of the F1 Scores on the validation sets of the clients for each step of the experiment. 

As we can observe in the figure, \ac{ourtool} produces high performance across various attack combinations. Regardless of the attack types used, \ac{ourtool} consistently achieves an F1 score above $0.96$, with a remarkably low standard deviation of less than $0.025$. The results obtained demonstrate the adaptability of \ac{ourtool} in adjusting its learning strategy in response to changes in data distribution and imbalance ratio. This adaptability extends to various scenarios, including those where only a few attacks are present, as well as instances involving out-of-distribution features, such as the two TCP-based attacks. 

However, in scenarios where only two attacks are present, we observe a slight drop in performance. This can be attributed to the adaptive selection of clients, which sometimes allocates computation to one client while leaving the other with none. Particularly when the feature distribution of the two attacks is significantly different, this dynamic assignment of computation may cause the F1 score of the global model to fluctuate between the two attack types, without any notable improvement in the average F1 score. Consequently, there are instances where the \ac{fl} process halts prematurely due to the ``patience'' mechanism, before achieving the expected average accuracy. In corner cases like this, it may be necessary to implement more sophisticated stopping strategies, as previously mentioned in Section \ref{sec:evaluation_discussion}.    

\subsection{Scalability analysis}
The previous sections detail experiments conducted in scenarios with federations of either $13$ or $50$ clients. However, these experiments do not provide insight into \ac{ourtool}'s stability when handling larger federations. To address this, we assess \ac{ourtool}'s performance across a range of increasing federation sizes, from $13$ to $90$ clients, measuring key metrics such as F1 Score and time required to reach convergence (inclusive of the $25$ rounds of patience).

To create federations of increasing sizes, we have progressively augmented the original set of $13$ clients with new clients generated by selecting two local datasets from the original set in various combinations. This process can produce up to $78$ new clients, each with a distinct local dataset, as calculated by the formula $n!/(k!(n-k)!)$ with $n=13$ and $k=2$, for a maximum federation size of $13+78=91$ clients.

\begin{figure}[!h]
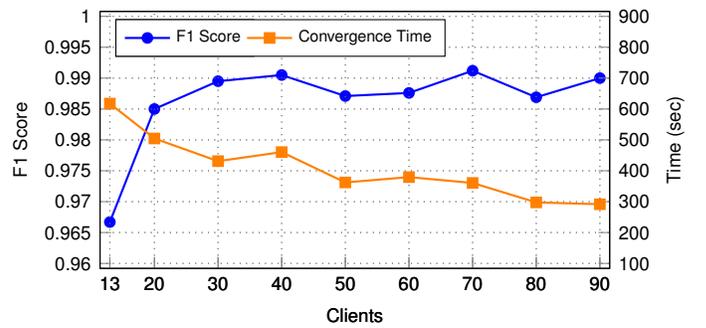

	\centering
	\chartScalabilityFLAD
	\caption{F1 Score and convergence time as functions of federation size.}
	\label{fig:flad_scalability}
\end{figure}

The experiment has been executed ten times for each federation size, with random subsets of clients selected at each iteration. The resulting average trends of F1 Score and convergence time are displayed in Figure \ref{fig:flad_scalability}.

The Figure clearly illustrates that \ac{ourtool} exhibits consistent stability in performance as the number of clients increases. Furthermore, these results validate our initial assumption that the one-to-one mapping between clients and attack types presents the most challenging scenario for the convergence of the \ac{fl} process. The data reveals the lowest F1 Score and the longest average convergence time during our experiments, supporting this conclusion.

Please keep in mind that the F1 Score for the global model is calculated by averaging the F1 Scores on all clients' validation sets. Therefore, when the number of clients is small, a low score on a few validation sets can have a significant impact on the overall average. In the scenario with $13$ clients, we observe an average F1 Score ranging between $0.90$ and $0.97$ on the validation sets of seven clients, those with WebDDoS (the worst with $0.90$), LDAP, Portmap, UDPLag, NTP, TFTP and NetBIOS attack data. In scenarios with more clients, where these attacks are present in multiple local datasets (although mixed with other attacks), \ac{ourtool} learns all attacks more accurately, although it consistently produces an F1 Score below $0.92$ on the WebDDoS attack.

The convergence time of the \ac{fl} process is greatly impacted by the global model's ability to perform well on the two TCP-based attacks. By adding new clients as described earlier in this section, we can potentially include more local datasets that contain TCP-based attack data, even if mixed with other types of attacks. This allows the global model to learn more quickly how to correctly classify the TCP-based attacks, which have proven to be critical for the convergence of the \ac{fl} process (see Section \ref{sec:convergence_analysis}).

%% file: discussion.tex
\revised{
\section{Discussion}\label{sec:discussion}
As previously elaborated in Section \ref{sec:threat_model}, the \ac{fl} process can potentially face security threats from malevolent clients and servers. These entities may attempt to exploit their positions within the federation to manipulate the classification performance of the global model or gain insight into the confidential data of other participants.

There are two main differences between \ac{ourtool} and \ac{fvg} approaches: (1) \ac{ourtool} involves clients sharing their accuracy metrics with the server, whereas \ac{fvg} requires the sharing of local training set sizes, and (ii) with \ac{ourtool} no test data is required at the server site to assess the quality of the global model. In situations where malicious clients seek to compromise the global model, they can employ conventional techniques such as model poisoning, label flipping, etc. \cite{xia2023poisoning} to manipulate the model's performance. This manipulation may result in the model missing certain types of \ac{ddos} attacks or other forms of intrusions. This is a general problem of \ac{fl} that has been tackled in previous studies \cite{yin2022trusted,qu2022blockchain,zhang2022secfednids,lai2023two,lycklama2023rofl}. To further exacerbate the challenges associated with \ac{fl}, a malicious client can employ different strategies. One approach involves transmitting fake information on the number of samples to the server ($n_k$ in Equation \ref{eq:fed_avg}), effectively assigning more weight to its manipulated contributions (\ac{fvg}). Alternatively, the client can deliberately provide a lower accuracy value, leading to an extended allocation of training rounds and epochs (\ac{ourtool}). Regarding the latter consideration, it is essential to recognise that a malicious client can independently determine the learning rate or the number of training epochs \cite{bagdasaryan2020backdoor}, irrespective of whether the \ac{fl} process employs \ac{ourtool} or \ac{fvg}. 
However, in the case of \ac{ourtool}, the client can strategically exploit the accuracy value to secure more training rounds than actually required, thereby influencing the global model's weights in favour of its malicious objectives (e.g., classifying a \ac{ddos} type as benign). In contrast, with \ac{fvg}, this manipulation is not feasible, as client selection is always random.

Clients of \ac{fl} may face the threat of reconstruction attacks perpetrated by a malicious server, which can exploit information on global model architecture, clients' gradients and other metadata to infer details of the original clients' training data \cite{truong2021privacy, na2022closing,10.5555/3495724.3497145}. This inherent vulnerability within \ac{fl} can become more pronounced in implementations where supplementary information is disclosed to the server. Examples of such information include the count of local training samples (as seen in \ac{fvg}) or the accuracy of the global model on the local validation sets (as in the case of \ac{ourtool}). 
To mitigate this vulnerability, various techniques have been recently proposed in the scientific literature, starting from differential privacy \cite{shen2022performance,hu2022federated}, which consists of adding noise to distort the shared parameters, to a novel approach based on obscuring the clients' gradients via fragmentation \cite{na2022closing}.  

In this study, our focus has been on addressing the challenges associated with achieving convergence in the \ac{fl} process within network intrusion detection scenarios. We recognise that issues of potential malicious clients and servers are crucial factors to consider when implementing an \ac{fl} framework. However, we acknowledge that these aspects fall outside the scope of our current work. 
Nevertheless, we consider them as opportunities for further investigation and exploration in the future.
}

%% file: conclusions.tex
\section{Conclusions}\label{sec:conclusions}
The main challenge in adopting \ac{fl} techniques in cybersecurity is assessing the performance of the global model on those attacks whose feature distributions are only known by clients. In this paper, we have presented \ac{ourtool}, an adaptive \ac{fl} approach for training feed-forward neural networks for \ac{ddos} attack detection, that implements a mechanism to monitor the classification accuracy of the global model on the clients' validations sets, without requiring any exchange of data. Thanks to this mechanism, \ac{ourtool} can estimate the performance of the aggregated model and dynamically tune the \ac{fl} process by assigning more computation to those clients whose attack profiles are harder to learn. \ac{ourtool} has been proven to significantly reduce convergence time while also enhancing classification accuracy when compared to current state-of-the-art \ac{fl} solutions.

We have validated \ac{ourtool} using an unbalanced dataset of non-\ac{iid} \ac{ddos} attacks. \revised{However, we see the potential of the FLAD's approach in other application domains where clients are expected to contribute with brand new data classes, whose profiles are not available to the server for the assessment of the global model. Although outside the scope of this work, we believe that an interesting research direction could be exploring the adaptability of FLAD to generic Network \acp{ids} in the presence of unknown network attack types, its relevance to host-based IDSs in contexts with zero-day vulnerabilities exploited to compromise computing infrastructure, and its potential portability to other domains such as image classification, where some image classes may be exclusively available in the local datasets of a subset of clients.}

%% file: ack.tex
\section*{Acknowledgement}
This work was partially supported by the European Union's Horizon Europe Programme under grant agreement No 101070473 (project FLUIDOS). 